\documentclass[reprint,superscriptaddress,amsmath,amssymb,longbibliography,twocolumn]{revtex4-2}

\usepackage{amsmath}
\usepackage{mathtools}
\usepackage{amssymb}
\usepackage{graphicx}
\usepackage{bm}
\usepackage[colorlinks, linkcolor=blue, citecolor=blue, urlcolor=blue,breaklinks=true]{hyperref}
\usepackage{color,dsfont,multirow,booktabs}
\usepackage{braket}
\usepackage{tabularx}
\usepackage{lipsum,booktabs}
\usepackage{ulem}

\newcommand\av[1]{\langle #1\rangle}

\newcommand\dg[1]{#1^{\dagger}}

\newcommand\numberthis{\addtocounter{equation}{1}\tag{\theequation}}

\begin{document}

\date{\today}

\title{Nonreciprocal Quantum Batteries}

\author{B. Ahmadi}
\email{borhan.ahmadi@ug.edu.pl}
\address{International Centre for Theory of Quantum Technologies, University of Gdansk, Jana Bażyńskiego 1A, 80-309 Gdansk, Poland}
\author{P. Mazurek}
\email{pawel.mazurek@ug.edu.pl}
\address{International Centre for Theory of Quantum Technologies, University of Gdansk, Jana Bażyńskiego 1A, 80-309 Gdansk, Poland}
\author{P. Horodecki}
\address{International Centre for Theory of Quantum Technologies, University of Gdansk, Jana Bażyńskiego 1A, 80-309 Gdansk, Poland}
\author{S. Barzanjeh}
\email{shabir.barzanjeh@ucalgary.ca}
\address{Department of Physics and Astronomy, University of Calgary, Calgary, AB T2N 1N4 Canada}

\begin{abstract}

Nonreciprocity, arising from the breaking of time-reversal symmetry, has become a fundamental tool in diverse quantum technology applications. It enables directional flow of signals and efficient noise suppression, constituting a key element in the architecture of current quantum information and computing systems.  Here we explore its potential in optimizing the charging dynamics of a quantum battery. By introducing nonreciprocity through reservoir engineering during the charging process, we induce a directed energy flow from the quantum charger to the battery, resulting in a substantial increase in energy accumulation. Despite local dissipation, the nonreciprocal approach demonstrates a fourfold increase in battery energy compared to conventional charger-battery systems. We demonstrate that employing a shared reservoir can establish an optimal condition where nonreciprocity enhances charging efficiency and elevates energy storage in the battery. This effect is observed in the stationary limit and remains applicable even in overdamped coupling regimes, eliminating the need for precise temporal control over evolution parameters. Our result can be extended to a chiral network of quantum nodes, serving as a multi-cell quantum battery system to enhance storage capacity. The proposed approach is straightforward to implement using current state-of-the-art quantum circuits, both in photonics and superconducting quantum systems. In a broader context, the concept of nonreciprocal charging has significant implications for sensing, energy capture, and storage technologies or studying quantum thermodynamics. 

\end{abstract}

\maketitle

\section{Introduction}

Nonreciprocal devices play an important role in optimizing system performance by suppressing undesired signal paths, noises, and spurious modes \cite{Jalas2013}. The discreet foundation of nonreciprocity lies in breaking time-reversal symmetry, a fundamental principle governing electromagnetic wave behavior \cite{Lodahl2017, PhysRevApplied.10.047001}. The intentional breaking of time-reversal symmetry emerges as a strategy in engineering nonreciprocal devices, providing precise control over signal pathways and improving the efficiency of quantum devices. Nonreciprocity not only facilitates selective signal transmission in one direction while hindering the opposite to prevent interference but also stimulates exploration into inventive quantum device functionalities. In circuit quantum electrodynamics \cite{Auld1959, pozar} and nanophotonics \cite{Li2014, PhysRevLett.109.033901, PhysRevLett.108.153901, Manipatruni2009} quantum communications, circulators serve as essential components, operating as single-port couplers or isolators. Their critical role lies in shielding vulnerable quantum states of cavities and qubits against electromagnetic noises and back reflection of intense signals/pumps.

In recent years, there has been a growing interest in both theoretical exploration and experimental application of nonreciprocity in a wide range of systems, spanning from microwave quantum circuits~\cite{Kamal2011,Abdo2013,Estep2014,Viola2014,
Sliwa2015,Kerckhoff2015,Lecocq2017,
Mahoney2017, Chapman1, Chapman2} to photonic devices \cite{Manipatruni2009, Bi2011, KangM.2011} and opto-electromechanical systems ~\cite{Hafezi2012a,Metelmann2015,Xu2015, Shen2016,Ruesink2016,Fang2017, Toth2017, Barzanjeh2017}. This interest is directed towards the design of efficient optical and microwave isolators and the investigation of various manifestations of nonreciprocity.
However, the importance of nonreciprocity goes beyond merely suppressing unwanted signals. For instance, nonreciprocal devices can be utilized to regulate the flow of thermal noise, enabling the realization of a thermal rectifier in nano-scale quantum devices \cite{PhysRevLett.120.060601}. In this paper, we study a novel aspect of nonreciprocal interactions: the role of nonreciprocity in facilitating efficient energy transfer and storage between quantum systems. We show that this concept has the potential for the implementation of nonreciprocal quantum batteries. 

The exploration of utilizing quantum systems as quantum batteries for capturing and storing energy has recently gained significant attention \cite{ Campaioli2023, PhysRevE.87.042123, PhysRevLett.118.150601, PhysRevB.99.205437, binder2015quantacell, farina2019charger, PhysRevLett.111.240401, PhysRevLett.120.117702, eabk3160, Barra, Bhattacharjee2021, PhysRevApplied.14.024092}. The primary aim of quantum batteries is to enhance the efficiency of the energy storage and charging power \cite{PhysRevE.87.042123, PhysRevLett.118.150601, PhysRevLett.120.117702, Barra, PhysRevApplied.14.024092, binder2015quantacell}, utilizing quantum resources such as entanglement \cite{PhysRevLett.120.117702,PhysRevLett.122.047702,PhysRevResearch.5.013155, PhysRevLett.120.117702}, quantum optimal control \cite{rodriguez2022optimal,PhysRevA.107.032218}, and quantum catalysis processes \cite{PhysRevA.107.042419}. Using interconnected quantum nodes/units offers the potential to design multi-cell quantum batteries, enhancing the capacity of energy storage. However, attempting to establish such a system solely through coherent interaction between cells is impractical. As with any other quantum systems, the challenges of decoherence and losses present a potential threat to preserving the essential quantum properties necessary for constructing reliable quantum batteries, as demonstrated experimentally \cite{Quach, Hu_2022, PhysRevA.106.042601,QUACH20232195}.

Building on the latest developments in the theory of engineered nonreciprocity, we introduce a novel approach designed to maximize energy storage in a quantum battery while minimizing energy dissipation to the surrounding environment during the charging process. Our method utilizes reservoir engineering \cite{PhysRevLett.77.4728, PhysRevLett.112.133904, PhysRevX.5.021025}, where a carefully designed dissipative environment facilitates efficient energy transfer between the modes of interest. Incorporating a dissipative reservoir, like an auxiliary waveguide, initiates an effective dissipative interaction between the charger and the battery. To create a nonreciprocal energy flow, we utilize an interference-like process, carefully balancing the induced dissipative interaction with its coherent counterpart. This nonreciprocal energy transfer significantly boosts the accumulation of energy in the quantum battery, enhancing energy storage capabilities and reducing energy back-flow to the charger. Our nonreciprocal battery is feasible for implementation using current experimental capabilities, whether in opto-electromechanical systems ~\cite{Toth2017, Barzanjeh2017}, superconducting circuits \cite{PhysRevApplied.4.034002, PhysRevX.7.041043}, or magnonic systems \cite{PhysRevLett.123.127202, 2303.04358}.

To explore the dynamics of the system, we model both the quantum charger and quantum battery as quantum harmonic oscillators experiencing energy dissipation into the surroundings. The quantum charger receives energy from a pump field, serving as the input energy source. We thoroughly investigate the behavior and characteristics of the proposed nonreciprocal system, examining the impact of reservoir engineering on its overall performance. Our results have the potential to contribute to the advancement of understanding and engineering nonreciprocal energy transfer in quantum systems. Specifically, this opens up a new avenue for enhancing energy storage efficiency in micro and nano-scaled quantum devices for future applications. 
\begin{figure}[t]
\centering
\includegraphics[width=8.5cm]{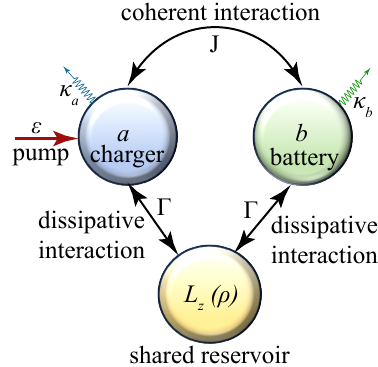}
\caption{Schematic representation of the nonreciprocal quantum battery system. A quantum charger $a$ coherently interacts with a quantum battery $b$ with a coupling rate $J$. The system's energy is supplied by an external pump with frequency $\omega_L$ and amplitude $\mathcal{E}$. Both the charger and the battery simultaneously couple to a shared reservoir with a rate $\Gamma$, establishing an effective dissipative interaction between them. The nonreciprocity is achieved by balancing the two interaction paths, specifically $\Gamma=-iJ/2$. In this configuration, energy directly flows from the charger to the battery, while the nonreciprocal condition effectively suppresses energy backflow. Here, $\kappa_a$ and $\kappa_b$ describe the local damping rates of each mode.}
\label{Fig1}
\end{figure}

Figure \ref{Fig1} illustrates the system's schematic, where a harmonic oscillator serves as the charger with a resonance frequency of $\omega_a$ and local damping rate $\kappa_a$. This charger interacts with another mode acting as the battery, which has a frequency of $\omega_b$ and a damping rate $\kappa_b$. The interaction between the charger and the battery is established through a coherent coupling with a rate of $J$. A classical drive field with a frequency of $\omega_L$ and amplitude $\mathcal{E}$ is utilized to provide energy to the charger. This energy then will be used to charge the battery. The Hamiltonian describing this driven bipartite system can be written as $(\hbar=1)$
\begin{equation}\label{Hamiltonian0}
    H=\omega_aa^\dagger a + \omega_bb^\dagger b + (J a^\dagger b + J^* b^\dagger a) + \mathcal{E} (e^{i\omega_L t}a + e^{-i\omega_L t}a^\dagger),
\end{equation}
where $a$ and $b$ are the annihilation operators of the charger and the battery, respectively.
For simplicity, we assume that both the charger and the battery have the same resonance frequency, $\omega=\omega_a=\omega_b$. 

Until now, the model remains similar to the charger-battery model previously studied in other works  \cite{PhysRevLett.120.117702, Andolina2018,farina2019charger}, where the reciprocal interaction between the charger and battery results in the exchange of energy between them. To achieve nonreciprocity, we additionally consider the existence of a dissipative interaction between the charger and the battery modes. This interaction occurs when both modes are coupled to a common (shared) nonlocal reservoir, characterized by a coupling rate $\Gamma$. By adiabatically eliminating the reservoir, an effective dissipative coupling between the charger and battery is established \cite{PhysRevX.5.021025}. In our analysis, we assume that the reservoir is Markovian and thus, we employ the standard master equation to describe the dynamics of the system $\dot{\rho}=-i[H,\rho] + \sum_{i=a,b}\kappa_i\mathcal{L}_i[\rho] + \Gamma\mathcal{L}_z[\rho],$ where $\mathcal{L}_c[\rho]=c\rho c^\dagger-\frac{1}{2}\{c^\dagger c,\rho\}$ represents the dissipative super-operator resulting from the coupling to the shared reservoir. Here, $z=p_a a+p_b b$ with $p_{a}$ and $p_b$ describe the coupling of the charger and battery, respectively, to the shared reservoir \cite{PhysRevX.5.021025}. For simplicity, we are considering a Markovian reservoir, but the nonreciprocity can be implemented in non-Markovian regimes as well. The first term of the master equation describes the coherent coupling between the charger and battery, while the second term describes the local damping of each mode to their bath with rate $\kappa_{a/b}$. The last term represents the dissipation of both modes into the shared reservoir, resulting in dissipative coupling between the two modes. This coupling can be realized by treating the shared bath as a damped cavity, waveguide, or transmission line \cite{Barzanjeh2017, PhysRevLett.123.127202, 2303.04358}. 

In the Schr\"odinger picture, the evolution of the operators is governed by the following equations
\begin{align*}\label{Ava}
    \frac{d\av{a}}{dt}&=-\Big(\frac{\Lambda_a}{2} +i\omega\Big)\av{a}-i\Big(J+i\mu\frac{\Gamma}{2}\Big)\av{b} -ie^{-i\omega_{L}t}\mathcal{E},\numberthis\\
\frac{d\av{b}}{dt}&=-\Big(\frac{\Lambda_b}{2} +i\omega\Big)\av{b}-i\Big(J^*+i\mu^*\frac{\Gamma}{2}\Big)\av{a}. \nonumber
\end{align*}
However, the energy stored in the battery is given by $E_B=\mathrm{Tr}_{a}[\rho H]=\omega_b \langle b^\dagger b\rangle$ where $\text {Tr}_{a} [..]$ is the partial trace over the charger's degree of freedom. Thus, calculating the energy of the system requires determining the second moments of the operators
\begin{align*}\label{Avada}
    \frac{d\av{\dg{a}a}}{dt}&=-\Lambda_a \av{\dg{a}a}-2\text{Re}\Big\{i\Big(J+i\mu\frac{\Gamma}{2}\Big)\av{\dg{a}b}\Big\} \\\nonumber
    &- 2\text{Im}\Big\{e^{i\omega_{L}t}\mathcal{E}\av{a}\Big\},\numberthis\\
        \frac{d\av{\dg{b}b}}{dt}&=-\Lambda_b \av{\dg{b}b}+2\text{Re}\Big\{i\Big(J-i\mu\frac{\Gamma}{2}\Big)\av{\dg{a}b}\Big\},\nonumber
\end{align*}
which explains that the population of each mode is directly influenced by $\av{\dg{a}b}$, indicating the excitation hopping between the two modes, expressed by
\begin{align*}\label{Avadb}
    \frac{d\av{\dg{a}b}}{dt}&=-\Big(\frac{\Lambda_a+\Lambda_b}{2}\Big)\av{\dg{a}b} - i\Big(J^*+i\mu^*\frac{\Gamma}{2}\Big)\av{\dg{a}a}\\\nonumber
    &+ i\Big(J^*-i\mu^*\frac{\Gamma}
{2}\Big)\av{\dg{b}b} + ie^{i\omega_{L}t}{\mathcal{E}}\av{b},\numberthis
\end{align*}
where $\mu=-p_{b}p^{*}_{a}$, $\Gamma_i=\Gamma|p_{i}|^{2}$, and $\Lambda_{i}=\Gamma_{i}+\kappa_{i}$ with $i=a,b$. As long as both $p_{a}$ and $p_{b}$ are nonzero, we can always re-scale them such that $|\mu| = 1$, as the remaining factor is absorbed into $\Gamma$.

The nonreciprocity now can be achieved by balancing the dissipative and coherent interaction rates. In Eqs. (\ref{Ava})--(\ref{Avadb}), by selecting specific values for the coupling $ J=-i\mu\frac{\Gamma}{2}$, 
the behavior of $\av{a}$ and $\av{\dg{a}a}$ remains entirely unaffected by the presence of the battery which implies a unidirectional flow of energy from the charger to the battery, without any energy backflow. This imposes nonreciprocal conditions and ensures an effective accumulation of energy in the battery.
Solving the complete set of Eqs. (\ref{Ava})--(\ref{Avadb}), we obtain the stored energy in the battery in the nonreciprocal regime
\begin{eqnarray}\label{AllTimeEnergyBNonReciprocal}\nonumber
     E_{B}^{\text{nr}}(t)&=&\frac{16\omega\Gamma^2\mathcal{E}^2}{\Lambda_a^2 \Lambda_b^2}\Big(1 + \Big[\frac{\Lambda_a e^{-\frac{1}{2}\Lambda_{b} t}-\Lambda_b e^{-\frac{1}{2}\Lambda_{a}t} }{(\Lambda_a-\Lambda_b)^2}\Big]\\
     &\times&\Big[(\Lambda_{a}e^{-\frac{1}{2}\Lambda_b t} - \Lambda_{b}e^{-\frac{1}{2}\Lambda_a t}) - 2(\Lambda_{a}-\Lambda_b)\Big]\Big),
\end{eqnarray}
where we considered the drive to be in resonance with each local mode $\delta=\omega_L-\omega=0$, see Appendix \ref{SM1} for the derivation and proof of optimality of the resonant drive. In the limit of symmetric damping rates, i.e., $\Lambda=\Lambda_{a}=\Lambda_{b}$ (which requires $\Gamma=\Gamma_a=\Gamma_b$ and $\kappa_a=\kappa_b$), Eq. (\ref{AllTimeEnergyBNonReciprocal}) can be simplified to
\begin{equation}\label{nonreci}
E_{B}^{\text{nr}}(t)=\frac{16\omega\Gamma^2\mathcal{E}^2}{\Lambda^4} \Big[1+e^{-\Lambda t/2}(e^{-\Lambda t/2}-2)\Big].
\end{equation}
where for $t \rightarrow \infty$, this function reaches its steady state value $E_{B}^{\text{nr}}(\infty)=\frac{16\omega\Gamma^2\mathcal{E}^2}{\Lambda^4}$. 

\begin{figure}[t]
\centering
\includegraphics[width=\linewidth]{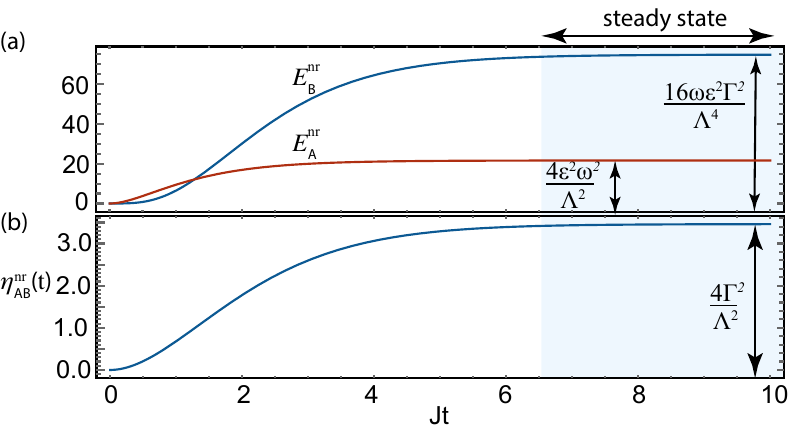}
\caption{(a) The total energy of the charger $E_A^{\text{nr}}$ and battery $E_B^{\text{nr}}$ plotted against the scaled time $Jt$ in the nonreciprocal regime $|J|=|\Gamma/2|$. This plot illustrates that as the system approaches a steady state, the energy of the battery surpasses that of the charger. (b) The ratio between the energies of the battery and charger, $\eta_{AB}^{\text{nr}}(t)=E_B^{\text{nr}}/E_A^{\text{nr}}$. In the steady-state limit, this ratio converges to $\mathcal{C}_d=4\Gamma^2/\Lambda^2$. In both figures, we consider $\Gamma=\Gamma_a=\Gamma_b=0.04\,\omega$, $\mathcal{E}=0.1\,\omega$, $\kappa_a=\kappa_b=0.003\,\omega$, and $J=\Gamma/2$.  }
\label{Fig2}
\end{figure}

We can also calculate the energy of the charger under nonreciprocal condition
\begin{equation}\label{AllTimeEnergy
ANonReciprocal}
    E_A^{\text{nr}}(t):=\omega\langle a^{\dagger}a\rangle = \frac{4\omega \mathcal{E}^2 \left(1-e^{-\frac{1}{2} \Lambda t}\right)^2}{\Lambda^2},
\end{equation}
with steady state value $E_A^{\text{nr}} (\infty)=\frac{4\omega \mathcal{E}^2}{\Lambda^2}$. In Fig. \ref{Fig2}a, the energies of the battery $E^{\text{nr}}_{B}$ and charger $E^{\text{nr}}_{A}$ are plotted against the scaled time $Jt$ in the nonreciprocal regime. Initially, the energy stored in the battery is lower than that in the charger, but it progressively increases and reaches its steady-state value. 

The ratio of the energies of the charger and battery $\eta_{AB}^{\text {nr}}(t)=\frac{ E^{\text{nr}}_{B}(t)}{ E^{\text{nr}}_{A}(t)}$ provides additional insights into the energy distribution between the two modes. In the steady-state limit, we find that $\eta_{AB}^{\text nr}(\infty)=\mathcal{C}_{d}$, where $\mathcal{C}_{d}=4\Gamma^2/\Lambda^2$ represents dissipative cooperativity between the charger and the battery through the shared reservoir. Efficient energy storage in the battery in the nonreciprocal scenario necessitates $\mathcal{C}_{d}>1$ or, equivalently, $\Gamma>\kappa$, where $\kappa=\kappa_a=\kappa_b$. Therefore, as long as the coupling of both the charger and the battery to the shared reservoir surpasses the local damping of each mode, the energy of the battery exceeds that of the charger. Figure \ref{Fig2}b illustrates the parameter $\eta_{AB}^{\text {nr}}(t)$ versus the scaled time $Jt$. This ratio surpasses 1 immediately after the interaction is turned on and reaches its steady-state value set by $\mathcal{C}_d$.

The efficiency of the nonreciprocal charging method can be examined by comparing the battery's energy to that of a conventional reciprocal charger-battery system \cite{farina2019charger, PhysRevB.99.205437}. This comparison can be conducted without taking the shared reservoir into account by setting $\Gamma_{i}=\Gamma=0$ in Eqs. (\ref{Ava})--(\ref{Avadb}), resulting in 
\begin{equation}\
    E_{B} = \zeta \Big(1 -\alpha(t)e^{-\kappa_{ab}t/4}+\Big[\beta(t)-\frac{2\kappa_a\kappa_b(\mathcal{C}+1)}{\Delta^2}\Big]e^{-\kappa_{ab}t/2}\Big),
     \label{energyRec}
\end{equation}
with  $\zeta=\frac{4\omega \mathcal{E}^2\mathcal{C}}{\kappa_a\kappa_b(\mathcal{C}+1)^2}$ and 
\begin{align}\label{const}\nonumber
    \alpha(t)&= \frac{2}{\Delta}\Big(\Delta\cosh{\Delta t/4}+\kappa_{ab}\sinh{\Delta t/4}\Big),\\\nonumber
     \beta(t)&= \frac{(\Delta^2+\kappa_{ab}^2)\cosh{\Delta t/2}-2\kappa_{ab}\Delta \sinh{\Delta t/2}}{2\Delta^2},\numberthis
\end{align}
where $\Delta=\sqrt{-16|J|^2+(\kappa_{a}-\kappa_{b})^2}$, $\kappa_{ab}=\kappa_a+\kappa_b$, and $\mathcal{C}=4J^2/\kappa_a\kappa_b$ is the coherent interaction cooperativity between charger and battery. 

In the underdamped regime when $4|J|>\{\kappa_{a},\kappa_{b}\}$ or for $4|J|>\{|\kappa_{a}-\kappa_{b}|\}$, $\Delta$ is imaginary. In this case, the hyperbolic functions in Eqs. (\ref{const}) transform into oscillating terms, representing the energy exchange between the battery and charger. These oscillations experience exponential decay due to the local damping in each mode with rate $\kappa_{a/b}$, leading to a slow and inefficient charging process. When the excitation returns to the charger, a portion of it leaks into the charger's environment before reaching back to the battery. In the nonreciprocal scenario, the dynamics of the charger is entirely separate from the battery. This isolation eliminates time-dependent oscillations in the battery's energy, enabling a more efficient and faster charging process.
In the following section, we first explore the underdamped regime ($\Delta$ is imaginary in general), comparing energy storage in the battery under nonreciprocal and reciprocal conditions. Subsequently, our attention turns to real values of $\Delta$ in the overdamped regime. In both regimes, we identify an optimal condition wherein nonreciprocity consistently leads to efficient energy storage in the battery.

\begin{figure}[t]
\centering
\includegraphics[width=\linewidth]{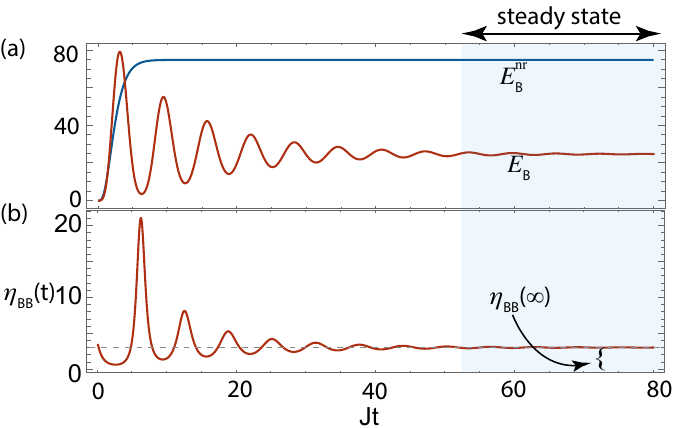}
\caption{(a) The energy of the battery in reciprocal $E_B (t)$ and nonreciprocal $E_B^{\text{nr}}(t)$ regimes plotted versus the scaled time $Jt$ for imaginary $\Delta$ when $4|J|>\{\kappa_{a},\kappa_{b}\}$ or $4|J|>\{|\kappa_{a}-\kappa_{b}|\}$. (b) The parameter $\eta_{BB}(t)=E_B^{\text{nr}} (t)/E_B (t)$ versus the scaled time $Jt$. In the steady-state limit, this ratio converges to $\eta_{BB}(\infty)$. In both figures, we consider $\Gamma=\Gamma_a=\Gamma_b=0.04\,\omega$, $\mathcal{E}=0.1\,\omega$, $\kappa_a=\kappa_b=0.003\,\omega$, and $J=\Gamma/2$, leading to $\Delta/\omega=0.08\,i$.  }
\label{Fig3}
\end{figure}

In Fig. \ref{Fig3}a, the comparison of the battery's energy with and without nonreciprocity is presented over the scaled time $Jt$ for $\kappa_a=\kappa_b$, where $\Delta$ is imaginary (underdamped i.e. $4|J|>\{\kappa_{a},\kappa_{b}\}$). Nonreciprocity exhibits superior results, approaching a steady-state regime, in contrast to the reciprocal charger-battery system. As for finite time dynamics, in the underdamped transient regime, while the energy of the battery in the nonreciprocal scenario is guaranteed to increase monotonically, the energy in the reciprocal scenario may oscillate, temporarily achieving values that are higher than stationary nonreciprocal values. 

 In general, however, the coupling of the charger and battery to the shared reservoir allows for the optimization of system parameters. In Appendix \ref{SM2}, we show that the optimal condition can be attained by setting $p_a=\xi{\kappa_{b}}$ and $p_b=1/\xi$ with $\xi=\sqrt\frac{\kappa_{a}}{\kappa_{b}}$.  Therefore for $\kappa_{b}<0.22\kappa_{a}\leq |J|$ the optimized nonreciprocal condition allows for achieving battery energy levels that are not accessible in reciprocal regime at any times of the evolution, the proof of which we present in Appendix \ref{SM3}.

The impact of nonreciprocity and its advantages becomes more apparent when examining the parameter $\eta_{BB}(t)=\frac{ E^{\text{nr}}_{B}(t)}{ E_B(t)}$, as seen in Fig. \ref{Fig3}b. This function oscillates and eventually reaches its steady state, set by
\begin{align}
    \eta_{BB}(\infty)=
 4\Big(\frac{1+\mathcal{C}}{(\sqrt{\mathcal{C}}+\xi)(\sqrt{\mathcal{C}}+\frac{1}{\xi})}\Big)^2,
\end{align}
where for a meaningful comparison, we use the nonreciprocity condition and set $|\Gamma|=2|J|$. When $\mathcal{C}>\xi^2$ or $\mathcal{C}<1/\xi^2$, the nonreciprocal regime yields higher stored energy in the battery compared to the reciprocal charger-battery regime. When $\kappa_a=\kappa_b$ or $\xi=1$, we observe $\eta\geq 1$. In such cases, for $\mathcal{C}\ll 1$ or $\mathcal{C}\gg 1$, the energy stored in the battery during the nonreciprocal regime is four times greater than that of a conventional charger-battery system $ \eta_{BB}(\infty) \approx 4$. 

\begin{figure}[t]
\centering
\includegraphics[width=\linewidth]{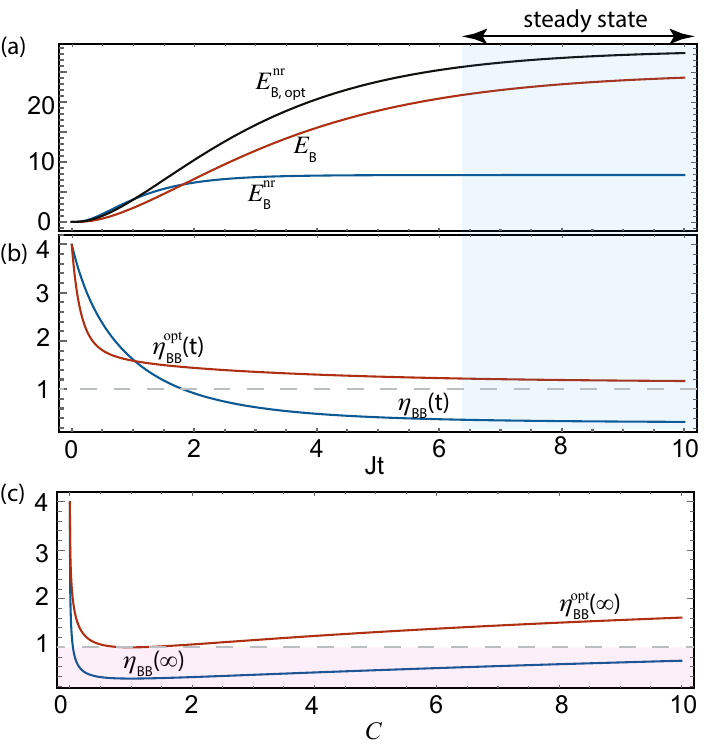}
\caption{
Comparing the energy of the battery in reciprocal $E_B (t)$, nonreciprocal $E_B^{\text{nr}} (t)$, and optimized nonreciprocal $E_{B,\text{opt}}^{\text{nr}} (t)$ regimes, presented against the scaled time $Jt$. (b) A comparison of the parameter $\eta_{BB}(t)$ and its optimized counterpart $\eta_{BB}^{\text{opt}}(t)$ over time. (c) The steady-state values of parameters $\eta_{BB} (\infty)$ and $\eta_{BB}^{\text{opt}}(\infty)$ versus cooperativity $\mathcal{C}$. For Figures (a) and (b), we assume $\Gamma=\Gamma_a=\Gamma_b=0.01\,\omega$, $\mathcal{E}=0.1\,\omega$, $\kappa_a=0.1\omega$, $\kappa_b=0.003\,\omega$, and $J=\Gamma/2$, corresponding to $\Delta\approx 0.095\, \omega$. In Figure (c), optimization conditions are applied with $\Gamma_a=\xi \Gamma$ and $\Gamma_b=\Gamma/\xi$, where $\xi=\sqrt{\kappa_a/\kappa_b}$.}
\label{Fig4}
\end{figure}

The result, however, can be extended to the overdamped regime or real values of $\Delta$ where $4|J|<\{\kappa_{a},\kappa_{b},|\kappa_a-\kappa_b|\}$. Figure \ref{Fig4}a, compares $E_B(t)$ and $E_B^{\text{nr}} (t)$ versus scaled time $Jt$ for $\kappa_a>\kappa_b$ and $4J^2<|\kappa_a-\kappa_b|^2$. As seen, the energy stored in the battery achieves its maximum without oscillations in both reciprocal and nonreciprocal cases. In this scenario, the reciprocal charger-battery approach yields superior results compared to the nonreciprocal charging mechanism. However, by performing optimization of system parameters and setting $\Gamma_a=\xi \Gamma$ and $\Gamma_b=\Gamma/\xi$, nonreciprocity can result in more efficient energy storage $E_{B, \text{opt}}^{\text{nr}}(t)$  in the battery, see Fig. \ref{Fig4}a. 

In Fig. \ref{Fig4}b, we plot the $\eta_{BB}(t)$ with and without optimization, confirming that the optimization leads to increased energy accumulation in the battery. In the steady state, we obtain
\begin{align}
    \eta_{BB}^{\text{opt}}(\infty)=\frac{\Gamma^2(4J^2+\kappa_a\kappa_b)^2}{J^2 (\Gamma+\sqrt{\kappa_a \kappa_b})^4},
\end{align}
and by setting $\Gamma=2J$ we get
\begin{align}
    \eta_{BB}^{\text{opt}}(\infty)=
 4\Big(\frac{1+\mathcal{C}}{(\sqrt{\mathcal{C}}+1)^2}\Big)^2,
\end{align}
in which for $\mathcal{C}\gg 1$ or $\mathcal{C}\ll 1$, $\eta_{BB}^{\text{opt}}\approx 4$, see Fig. \ref{Fig4}c. We note that this result is general, and as long as the optimal condition holds ($\Gamma_a=\xi \Gamma$ and $\Gamma_b=\Gamma/\xi$), nonreciprocity consistently outperforms the reciprocal charger-battery system.

In this paper, we theoretically proposed and investigated a nonreciprocal approach to enhance energy storage in a quantum battery. We demonstrated how breaking reciprocity via reservoir engineering enables unidirectional energy flow, reducing energy dissipation into the surroundings during the charging process, particularly advantageous for higher dissipation rates. The underlying physics relies on an interference-like phenomenon where coherent coupling between the charger and battery counteracts dissipative interaction. Here we did not discuss the work extracted from the battery or study ergotropy. The reason for that is when a quantum harmonic oscillator is linearly coupled to a bath at zero temperature, Markovian dynamics map a coherent state into another coherent state \cite{kossakowski1972quantum}. Assuming the initial state of the entire charger-battery system is a coherent state, it remains coherent and pure throughout the entire evolution, implying that all internal energy of the battery can be extracted as ergotropy.

Our proposed approach can be implemented using current state-of-the-art quantum photonic systems or microwave superconducting circuits.  Moreover, this system is scalable to a chain of batteries, forming a chiral network of quantum cells. The potential benefits include enhanced energy storage efficiency and capacity, offering a pathway for realistic quantum batteries in nano/micro-scaled quantum devices. In the broader context of advancing quantum technology with more complex systems and numerous qubits, quantum batteries stand out as potentially integral components. Their usage can extend to enabling reversible logic gates in quantum computing processors \cite{PhysRevX.11.021014}. Beyond that, our research introduces novel possibilities, such as leveraging nonreciprocity for the study of quantum thermodynamic phenomena and understanding energy flow in quantum devices. Additionally, it could spark the exploration of quantum batteries in naturally nonreciprocal materials or magneto-optic systems.

\subsection*{Acknowledgements}
We acknowledge support from the Foundation for Polish Science through IRAP project co-financed by EU within the Smart Growth Operational Program (contract no.2018/MAB/5). S.B. acknowledges funding by the Natural Sciences and Engineering Research Council of Canada (NSERC) through its Discovery Grant, funding and advisory support provided by Alberta Innovates through the Accelerating Innovations into CarE (AICE) -- Concepts Program, and support from Alberta Innovates and NSERC through Advance Grant.

\bibliography{References}

\begin{thebibliography}{61}%
\makeatletter
\providecommand \@ifxundefined [1]{%
 \@ifx{#1\undefined}
}%
\providecommand \@ifnum [1]{%
 \ifnum #1\expandafter \@firstoftwo
 \else \expandafter \@secondoftwo
 \fi
}%
\providecommand \@ifx [1]{%
 \ifx #1\expandafter \@firstoftwo
 \else \expandafter \@secondoftwo
 \fi
}%
\providecommand \natexlab [1]{#1}%
\providecommand \enquote  [1]{``#1''}%
\providecommand \bibnamefont  [1]{#1}%
\providecommand \bibfnamefont [1]{#1}%
\providecommand \citenamefont [1]{#1}%
\providecommand \href@noop [0]{\@secondoftwo}%
\providecommand \href [0]{\begingroup \@sanitize@url \@href}%
\providecommand \@href[1]{\@@startlink{#1}\@@href}%
\providecommand \@@href[1]{\endgroup#1\@@endlink}%
\providecommand \@sanitize@url [0]{\catcode `\\12\catcode `\$12\catcode `\&12\catcode `\#12\catcode `\^12\catcode `\_12\catcode `\%12\relax}%
\providecommand \@@startlink[1]{}%
\providecommand \@@endlink[0]{}%
\providecommand \url  [0]{\begingroup\@sanitize@url \@url }%
\providecommand \@url [1]{\endgroup\@href {#1}{\urlprefix }}%
\providecommand \urlprefix  [0]{URL }%
\providecommand \Eprint [0]{\href }%
\providecommand \doibase [0]{https://doi.org/}%
\providecommand \selectlanguage [0]{\@gobble}%
\providecommand \bibinfo  [0]{\@secondoftwo}%
\providecommand \bibfield  [0]{\@secondoftwo}%
\providecommand \translation [1]{[#1]}%
\providecommand \BibitemOpen [0]{}%
\providecommand \bibitemStop [0]{}%
\providecommand \bibitemNoStop [0]{.\EOS\space}%
\providecommand \EOS [0]{\spacefactor3000\relax}%
\providecommand \BibitemShut  [1]{\csname bibitem#1\endcsname}%
\let\auto@bib@innerbib\@empty
\bibitem [{\citenamefont {Jalas}\ \emph {et~al.}(2013)\citenamefont {Jalas}, \citenamefont {Petrov}, \citenamefont {Eich}, \citenamefont {Freude}, \citenamefont {Fan}, \citenamefont {Yu}, \citenamefont {Baets}, \citenamefont {Popovic}, \citenamefont {Melloni}, \citenamefont {Joannopoulos}, \citenamefont {Vanwolleghem}, \citenamefont {Doerr},\ and\ \citenamefont {Renner}}]{Jalas2013}%
  \BibitemOpen
  \bibfield  {author} {\bibinfo {author} {\bibfnamefont {D.}~\bibnamefont {Jalas}}, \bibinfo {author} {\bibfnamefont {A.}~\bibnamefont {Petrov}}, \bibinfo {author} {\bibfnamefont {M.}~\bibnamefont {Eich}}, \bibinfo {author} {\bibfnamefont {W.}~\bibnamefont {Freude}}, \bibinfo {author} {\bibfnamefont {S.}~\bibnamefont {Fan}}, \bibinfo {author} {\bibfnamefont {Z.}~\bibnamefont {Yu}}, \bibinfo {author} {\bibfnamefont {R.}~\bibnamefont {Baets}}, \bibinfo {author} {\bibfnamefont {M.}~\bibnamefont {Popovic}}, \bibinfo {author} {\bibfnamefont {A.}~\bibnamefont {Melloni}}, \bibinfo {author} {\bibfnamefont {J.~D.}\ \bibnamefont {Joannopoulos}}, \bibinfo {author} {\bibfnamefont {M.}~\bibnamefont {Vanwolleghem}}, \bibinfo {author} {\bibfnamefont {C.~R.}\ \bibnamefont {Doerr}},\ and\ \bibinfo {author} {\bibfnamefont {H.}~\bibnamefont {Renner}},\ }\bibfield  {title} {\bibinfo {title} {What is -- and what is not -- an optical isolator},\ }\href {http://dx.doi.org/10.1038/nphoton.2013.185} {\bibfield  {journal} {\bibinfo
  {journal} {Nat Photon}\ }\textbf {\bibinfo {volume} {7}},\ \bibinfo {pages} {579} (\bibinfo {year} {2013})}\BibitemShut {NoStop}%
\bibitem [{\citenamefont {Lodahl}\ \emph {et~al.}(2017)\citenamefont {Lodahl}, \citenamefont {Mahmoodian}, \citenamefont {Stobbe}, \citenamefont {Rauschenbeutel}, \citenamefont {Schneeweiss}, \citenamefont {Volz}, \citenamefont {Pichler},\ and\ \citenamefont {Zoller}}]{Lodahl2017}%
  \BibitemOpen
  \bibfield  {author} {\bibinfo {author} {\bibfnamefont {P.}~\bibnamefont {Lodahl}}, \bibinfo {author} {\bibfnamefont {S.}~\bibnamefont {Mahmoodian}}, \bibinfo {author} {\bibfnamefont {S.}~\bibnamefont {Stobbe}}, \bibinfo {author} {\bibfnamefont {A.}~\bibnamefont {Rauschenbeutel}}, \bibinfo {author} {\bibfnamefont {P.}~\bibnamefont {Schneeweiss}}, \bibinfo {author} {\bibfnamefont {J.}~\bibnamefont {Volz}}, \bibinfo {author} {\bibfnamefont {H.}~\bibnamefont {Pichler}},\ and\ \bibinfo {author} {\bibfnamefont {P.}~\bibnamefont {Zoller}},\ }\bibfield  {title} {\bibinfo {title} {Chiral quantum optics},\ }\href {http://dx.doi.org/10.1038/nature21037} {\bibfield  {journal} {\bibinfo  {journal} {Nature}\ }\textbf {\bibinfo {volume} {541}},\ \bibinfo {pages} {473} (\bibinfo {year} {2017})}\BibitemShut {NoStop}%
\bibitem [{\citenamefont {Caloz}\ \emph {et~al.}(2018)\citenamefont {Caloz}, \citenamefont {Al\`u}, \citenamefont {Tretyakov}, \citenamefont {Sounas}, \citenamefont {Achouri},\ and\ \citenamefont {Deck-L\'eger}}]{PhysRevApplied.10.047001}%
  \BibitemOpen
  \bibfield  {author} {\bibinfo {author} {\bibfnamefont {C.}~\bibnamefont {Caloz}}, \bibinfo {author} {\bibfnamefont {A.}~\bibnamefont {Al\`u}}, \bibinfo {author} {\bibfnamefont {S.}~\bibnamefont {Tretyakov}}, \bibinfo {author} {\bibfnamefont {D.}~\bibnamefont {Sounas}}, \bibinfo {author} {\bibfnamefont {K.}~\bibnamefont {Achouri}},\ and\ \bibinfo {author} {\bibfnamefont {Z.-L.}\ \bibnamefont {Deck-L\'eger}},\ }\bibfield  {title} {\bibinfo {title} {Electromagnetic nonreciprocity},\ }\href {https://doi.org/10.1103/PhysRevApplied.10.047001} {\bibfield  {journal} {\bibinfo  {journal} {Phys. Rev. Appl.}\ }\textbf {\bibinfo {volume} {10}},\ \bibinfo {pages} {047001} (\bibinfo {year} {2018})}\BibitemShut {NoStop}%
\bibitem [{\citenamefont {Auld}(1959)}]{Auld1959}%
  \BibitemOpen
  \bibfield  {author} {\bibinfo {author} {\bibfnamefont {B.~A.}\ \bibnamefont {Auld}},\ }\bibfield  {title} {\bibinfo {title} {The synthesis of symmetrical waveguide circulators},\ }\href {https://doi.org/10.1109/TMTT.1959.1124688} {\bibfield  {journal} {\bibinfo  {journal} {IRE Transactions on Microwave Theory and Techniques}\ }\textbf {\bibinfo {volume} {7}},\ \bibinfo {pages} {238} (\bibinfo {year} {1959})}\BibitemShut {NoStop}%
\bibitem [{\citenamefont {Pozar}(1993)}]{pozar}%
  \BibitemOpen
  \bibfield  {author} {\bibinfo {author} {\bibfnamefont {D.}~\bibnamefont {Pozar}},\ }\bibfield  {title} {\bibinfo {title} {Microwave engineering},\ }\href@noop {} {\bibfield  {journal} {\bibinfo  {journal} {Addison-Wesley Publishing Company}\ } (\bibinfo {year} {1993})}\BibitemShut {NoStop}%
\bibitem [{\citenamefont {Li}\ \emph {et~al.}(2014)\citenamefont {Li}, \citenamefont {Eggleton}, \citenamefont {Fang},\ and\ \citenamefont {Fan}}]{Li2014}%
  \BibitemOpen
  \bibfield  {author} {\bibinfo {author} {\bibfnamefont {E.}~\bibnamefont {Li}}, \bibinfo {author} {\bibfnamefont {B.~J.}\ \bibnamefont {Eggleton}}, \bibinfo {author} {\bibfnamefont {K.}~\bibnamefont {Fang}},\ and\ \bibinfo {author} {\bibfnamefont {S.}~\bibnamefont {Fan}},\ }\bibfield  {title} {\bibinfo {title} {Photonic aharonov--bohm effect in photon--phonon interactions},\ }\href {https://doi.org/10.1038/ncomms4225} {\bibfield  {journal} {\bibinfo  {journal} {Nature Communications}\ }\textbf {\bibinfo {volume} {5}},\ \bibinfo {pages} {3225} (\bibinfo {year} {2014})}\BibitemShut {NoStop}%
\bibitem [{\citenamefont {Lira}\ \emph {et~al.}(2012)\citenamefont {Lira}, \citenamefont {Yu}, \citenamefont {Fan},\ and\ \citenamefont {Lipson}}]{PhysRevLett.109.033901}%
  \BibitemOpen
  \bibfield  {author} {\bibinfo {author} {\bibfnamefont {H.}~\bibnamefont {Lira}}, \bibinfo {author} {\bibfnamefont {Z.}~\bibnamefont {Yu}}, \bibinfo {author} {\bibfnamefont {S.}~\bibnamefont {Fan}},\ and\ \bibinfo {author} {\bibfnamefont {M.}~\bibnamefont {Lipson}},\ }\bibfield  {title} {\bibinfo {title} {Electrically driven nonreciprocity induced by interband photonic transition on a silicon chip},\ }\href {https://doi.org/10.1103/PhysRevLett.109.033901} {\bibfield  {journal} {\bibinfo  {journal} {Phys. Rev. Lett.}\ }\textbf {\bibinfo {volume} {109}},\ \bibinfo {pages} {033901} (\bibinfo {year} {2012})}\BibitemShut {NoStop}%
\bibitem [{\citenamefont {Fang}\ \emph {et~al.}(2012)\citenamefont {Fang}, \citenamefont {Yu},\ and\ \citenamefont {Fan}}]{PhysRevLett.108.153901}%
  \BibitemOpen
  \bibfield  {author} {\bibinfo {author} {\bibfnamefont {K.}~\bibnamefont {Fang}}, \bibinfo {author} {\bibfnamefont {Z.}~\bibnamefont {Yu}},\ and\ \bibinfo {author} {\bibfnamefont {S.}~\bibnamefont {Fan}},\ }\bibfield  {title} {\bibinfo {title} {Photonic aharonov-bohm effect based on dynamic modulation},\ }\href {https://doi.org/10.1103/PhysRevLett.108.153901} {\bibfield  {journal} {\bibinfo  {journal} {Phys. Rev. Lett.}\ }\textbf {\bibinfo {volume} {108}},\ \bibinfo {pages} {153901} (\bibinfo {year} {2012})}\BibitemShut {NoStop}%
\bibitem [{\citenamefont {Manipatruni}\ \emph {et~al.}(2009)\citenamefont {Manipatruni}, \citenamefont {Robinson},\ and\ \citenamefont {Lipson}}]{Manipatruni2009}%
  \BibitemOpen
  \bibfield  {author} {\bibinfo {author} {\bibfnamefont {S.}~\bibnamefont {Manipatruni}}, \bibinfo {author} {\bibfnamefont {J.~T.}\ \bibnamefont {Robinson}},\ and\ \bibinfo {author} {\bibfnamefont {M.}~\bibnamefont {Lipson}},\ }\bibfield  {title} {\bibinfo {title} {Optical nonreciprocity in optomechanical structures},\ }\href {https://link.aps.org/doi/10.1103/PhysRevLett.102.213903} {\bibfield  {journal} {\bibinfo  {journal} {Phys. Rev. Lett.}\ }\textbf {\bibinfo {volume} {102}},\ \bibinfo {pages} {213903} (\bibinfo {year} {2009})}\BibitemShut {NoStop}%
\bibitem [{\citenamefont {Kamal}\ \emph {et~al.}(2011)\citenamefont {Kamal}, \citenamefont {Clarke},\ and\ \citenamefont {Devoret}}]{Kamal2011}%
  \BibitemOpen
  \bibfield  {author} {\bibinfo {author} {\bibfnamefont {A.}~\bibnamefont {Kamal}}, \bibinfo {author} {\bibfnamefont {J.}~\bibnamefont {Clarke}},\ and\ \bibinfo {author} {\bibfnamefont {M.~H.}\ \bibnamefont {Devoret}},\ }\bibfield  {title} {\bibinfo {title} {Noiseless non-reciprocity in a parametric active device},\ }\href {http://dx.doi.org/10.1038/nphys1893} {\bibfield  {journal} {\bibinfo  {journal} {Nat Phys}\ }\textbf {\bibinfo {volume} {7}},\ \bibinfo {pages} {311} (\bibinfo {year} {2011})}\BibitemShut {NoStop}%
\bibitem [{\citenamefont {Abdo}\ \emph {et~al.}(2013)\citenamefont {Abdo}, \citenamefont {Sliwa}, \citenamefont {Frunzio},\ and\ \citenamefont {Devoret}}]{Abdo2013}%
  \BibitemOpen
  \bibfield  {author} {\bibinfo {author} {\bibfnamefont {B.}~\bibnamefont {Abdo}}, \bibinfo {author} {\bibfnamefont {K.}~\bibnamefont {Sliwa}}, \bibinfo {author} {\bibfnamefont {L.}~\bibnamefont {Frunzio}},\ and\ \bibinfo {author} {\bibfnamefont {M.}~\bibnamefont {Devoret}},\ }\bibfield  {title} {\bibinfo {title} {Directional amplification with a {Josephson} circuit},\ }\href {https://link.aps.org/doi/10.1103/PhysRevX.3.031001} {\bibfield  {journal} {\bibinfo  {journal} {Phys. Rev. X}\ }\textbf {\bibinfo {volume} {3}},\ \bibinfo {pages} {031001} (\bibinfo {year} {2013})}\BibitemShut {NoStop}%
\bibitem [{\citenamefont {Estep}\ \emph {et~al.}(2014)\citenamefont {Estep}, \citenamefont {Sounas}, \citenamefont {Soric},\ and\ \citenamefont {Alu}}]{Estep2014}%
  \BibitemOpen
  \bibfield  {author} {\bibinfo {author} {\bibfnamefont {N.~A.}\ \bibnamefont {Estep}}, \bibinfo {author} {\bibfnamefont {D.~L.}\ \bibnamefont {Sounas}}, \bibinfo {author} {\bibfnamefont {J.}~\bibnamefont {Soric}},\ and\ \bibinfo {author} {\bibfnamefont {A.}~\bibnamefont {Alu}},\ }\bibfield  {title} {\bibinfo {title} {Magnetic-free non-reciprocity and isolation based on parametrically modulated coupled-resonator loops},\ }\href {http://dx.doi.org/10.1038/nphys3134} {\bibfield  {journal} {\bibinfo  {journal} {Nat Phys}\ }\textbf {\bibinfo {volume} {10}},\ \bibinfo {pages} {923} (\bibinfo {year} {2014})}\BibitemShut {NoStop}%
\bibitem [{\citenamefont {Viola}\ and\ \citenamefont {DiVincenzo}(2014)}]{Viola2014}%
  \BibitemOpen
  \bibfield  {author} {\bibinfo {author} {\bibfnamefont {G.}~\bibnamefont {Viola}}\ and\ \bibinfo {author} {\bibfnamefont {D.~P.}\ \bibnamefont {DiVincenzo}},\ }\bibfield  {title} {\bibinfo {title} {Hall effect gyrators and circulators},\ }\href {https://link.aps.org/doi/10.1103/PhysRevX.4.021019} {\bibfield  {journal} {\bibinfo  {journal} {Phys. Rev. X}\ }\textbf {\bibinfo {volume} {4}},\ \bibinfo {pages} {021019} (\bibinfo {year} {2014})}\BibitemShut {NoStop}%
\bibitem [{\citenamefont {Sliwa}\ \emph {et~al.}(2015)\citenamefont {Sliwa}, \citenamefont {Hatridge}, \citenamefont {Narla}, \citenamefont {Shankar}, \citenamefont {Frunzio}, \citenamefont {Schoelkopf},\ and\ \citenamefont {Devoret}}]{Sliwa2015}%
  \BibitemOpen
  \bibfield  {author} {\bibinfo {author} {\bibfnamefont {K.~M.}\ \bibnamefont {Sliwa}}, \bibinfo {author} {\bibfnamefont {M.}~\bibnamefont {Hatridge}}, \bibinfo {author} {\bibfnamefont {A.}~\bibnamefont {Narla}}, \bibinfo {author} {\bibfnamefont {S.}~\bibnamefont {Shankar}}, \bibinfo {author} {\bibfnamefont {L.}~\bibnamefont {Frunzio}}, \bibinfo {author} {\bibfnamefont {R.~J.}\ \bibnamefont {Schoelkopf}},\ and\ \bibinfo {author} {\bibfnamefont {M.~H.}\ \bibnamefont {Devoret}},\ }\bibfield  {title} {\bibinfo {title} {Reconfigurable josephson circulator/directional amplifier},\ }\href {https://link.aps.org/doi/10.1103/PhysRevX.5.041020} {\bibfield  {journal} {\bibinfo  {journal} {Phys. Rev. X}\ }\textbf {\bibinfo {volume} {5}},\ \bibinfo {pages} {041020} (\bibinfo {year} {2015})}\BibitemShut {NoStop}%
\bibitem [{\citenamefont {Kerckhoff}\ \emph {et~al.}(2015{\natexlab{a}})\citenamefont {Kerckhoff}, \citenamefont {Lalumière}, \citenamefont {Chapman}, \citenamefont {Blais},\ and\ \citenamefont {Lehnert}}]{Kerckhoff2015}%
  \BibitemOpen
  \bibfield  {author} {\bibinfo {author} {\bibfnamefont {J.}~\bibnamefont {Kerckhoff}}, \bibinfo {author} {\bibfnamefont {K.}~\bibnamefont {Lalumière}}, \bibinfo {author} {\bibfnamefont {B.~J.}\ \bibnamefont {Chapman}}, \bibinfo {author} {\bibfnamefont {A.}~\bibnamefont {Blais}},\ and\ \bibinfo {author} {\bibfnamefont {K.~W.}\ \bibnamefont {Lehnert}},\ }\bibfield  {title} {\bibinfo {title} {On-chip superconducting microwave circulator from synthetic rotation},\ }\href {https://link.aps.org/doi/10.1103/PhysRevApplied.4.034002} {\bibfield  {journal} {\bibinfo  {journal} {Phys. Rev. Applied}\ }\textbf {\bibinfo {volume} {4}},\ \bibinfo {pages} {034002} (\bibinfo {year} {2015}{\natexlab{a}})}\BibitemShut {NoStop}%
\bibitem [{\citenamefont {Lecocq}\ \emph {et~al.}(2017)\citenamefont {Lecocq}, \citenamefont {Ranzani}, \citenamefont {Peterson}, \citenamefont {Cicak}, \citenamefont {Simmonds}, \citenamefont {Teufel},\ and\ \citenamefont {Aumentado}}]{Lecocq2017}%
  \BibitemOpen
  \bibfield  {author} {\bibinfo {author} {\bibfnamefont {F.}~\bibnamefont {Lecocq}}, \bibinfo {author} {\bibfnamefont {L.}~\bibnamefont {Ranzani}}, \bibinfo {author} {\bibfnamefont {G.~A.}\ \bibnamefont {Peterson}}, \bibinfo {author} {\bibfnamefont {K.}~\bibnamefont {Cicak}}, \bibinfo {author} {\bibfnamefont {R.~W.}\ \bibnamefont {Simmonds}}, \bibinfo {author} {\bibfnamefont {J.~D.}\ \bibnamefont {Teufel}},\ and\ \bibinfo {author} {\bibfnamefont {J.}~\bibnamefont {Aumentado}},\ }\bibfield  {title} {\bibinfo {title} {Nonreciprocal microwave signal processing with a field-programmable josephson amplifier},\ }\href {https://doi.org/10.1103/PhysRevApplied.7.024028} {\bibfield  {journal} {\bibinfo  {journal} {Phys. Rev. Applied}\ }\textbf {\bibinfo {volume} {7}},\ \bibinfo {pages} {024028} (\bibinfo {year} {2017})}\BibitemShut {NoStop}%
\bibitem [{\citenamefont {Mahoney}\ \emph {et~al.}(2017)\citenamefont {Mahoney}, \citenamefont {Colless}, \citenamefont {Pauka}, \citenamefont {Hornibrook}, \citenamefont {Watson}, \citenamefont {Gardner}, \citenamefont {Manfra}, \citenamefont {Doherty},\ and\ \citenamefont {Reilly}}]{Mahoney2017}%
  \BibitemOpen
  \bibfield  {author} {\bibinfo {author} {\bibfnamefont {A.~C.}\ \bibnamefont {Mahoney}}, \bibinfo {author} {\bibfnamefont {J.~I.}\ \bibnamefont {Colless}}, \bibinfo {author} {\bibfnamefont {S.~J.}\ \bibnamefont {Pauka}}, \bibinfo {author} {\bibfnamefont {J.~M.}\ \bibnamefont {Hornibrook}}, \bibinfo {author} {\bibfnamefont {J.~D.}\ \bibnamefont {Watson}}, \bibinfo {author} {\bibfnamefont {G.~C.}\ \bibnamefont {Gardner}}, \bibinfo {author} {\bibfnamefont {M.~J.}\ \bibnamefont {Manfra}}, \bibinfo {author} {\bibfnamefont {A.~C.}\ \bibnamefont {Doherty}},\ and\ \bibinfo {author} {\bibfnamefont {D.~J.}\ \bibnamefont {Reilly}},\ }\bibfield  {title} {\bibinfo {title} {On-chip microwave quantum hall circulator},\ }\href {https://link.aps.org/doi/10.1103/PhysRevX.7.011007} {\bibfield  {journal} {\bibinfo  {journal} {Phys. Rev. X}\ }\textbf {\bibinfo {volume} {7}},\ \bibinfo {pages} {011007} (\bibinfo {year} {2017})}\BibitemShut {NoStop}%
\bibitem [{\citenamefont {Rosenthal}\ \emph {et~al.}(2017)\citenamefont {Rosenthal}, \citenamefont {Chapman}, \citenamefont {Higginbotham}, \citenamefont {Kerckhoff},\ and\ \citenamefont {Lehnert}}]{Chapman1}%
  \BibitemOpen
  \bibfield  {author} {\bibinfo {author} {\bibfnamefont {E.~I.}\ \bibnamefont {Rosenthal}}, \bibinfo {author} {\bibfnamefont {B.~J.}\ \bibnamefont {Chapman}}, \bibinfo {author} {\bibfnamefont {A.~P.}\ \bibnamefont {Higginbotham}}, \bibinfo {author} {\bibfnamefont {J.}~\bibnamefont {Kerckhoff}},\ and\ \bibinfo {author} {\bibfnamefont {K.~W.}\ \bibnamefont {Lehnert}},\ }\bibfield  {title} {\bibinfo {title} {Breaking lorentz reciprocity with frequency conversion and delay},\ }\href {https://doi.org/10.1103/PhysRevLett.119.147703} {\bibfield  {journal} {\bibinfo  {journal} {Phys. Rev. Lett.}\ }\textbf {\bibinfo {volume} {119}},\ \bibinfo {pages} {147703} (\bibinfo {year} {2017})}\BibitemShut {NoStop}%
\bibitem [{\citenamefont {Chapman}\ \emph {et~al.}(2017{\natexlab{a}})\citenamefont {Chapman}, \citenamefont {Rosenthal}, \citenamefont {Kerckhoff}, \citenamefont {Moores}, \citenamefont {Vale}, \citenamefont {Mates}, \citenamefont {Hilton}, \citenamefont {Lalumi\`ere}, \citenamefont {Blais},\ and\ \citenamefont {Lehnert}}]{Chapman2}%
  \BibitemOpen
  \bibfield  {author} {\bibinfo {author} {\bibfnamefont {B.~J.}\ \bibnamefont {Chapman}}, \bibinfo {author} {\bibfnamefont {E.~I.}\ \bibnamefont {Rosenthal}}, \bibinfo {author} {\bibfnamefont {J.}~\bibnamefont {Kerckhoff}}, \bibinfo {author} {\bibfnamefont {B.~A.}\ \bibnamefont {Moores}}, \bibinfo {author} {\bibfnamefont {L.~R.}\ \bibnamefont {Vale}}, \bibinfo {author} {\bibfnamefont {J.~A.~B.}\ \bibnamefont {Mates}}, \bibinfo {author} {\bibfnamefont {G.~C.}\ \bibnamefont {Hilton}}, \bibinfo {author} {\bibfnamefont {K.}~\bibnamefont {Lalumi\`ere}}, \bibinfo {author} {\bibfnamefont {A.}~\bibnamefont {Blais}},\ and\ \bibinfo {author} {\bibfnamefont {K.~W.}\ \bibnamefont {Lehnert}},\ }\bibfield  {title} {\bibinfo {title} {Widely tunable on-chip microwave circulator for superconducting quantum circuits},\ }\href {https://doi.org/10.1103/PhysRevX.7.041043} {\bibfield  {journal} {\bibinfo  {journal} {Phys. Rev. X}\ }\textbf {\bibinfo {volume} {7}},\ \bibinfo {pages} {041043} (\bibinfo {year}
  {2017}{\natexlab{a}})}\BibitemShut {NoStop}%
\bibitem [{\citenamefont {Bi}\ \emph {et~al.}(2011)\citenamefont {Bi}, \citenamefont {Hu}, \citenamefont {Jiang}, \citenamefont {Kim}, \citenamefont {Dionne}, \citenamefont {Kimerling},\ and\ \citenamefont {RossC.}}]{Bi2011}%
  \BibitemOpen
  \bibfield  {author} {\bibinfo {author} {\bibfnamefont {L.}~\bibnamefont {Bi}}, \bibinfo {author} {\bibfnamefont {J.}~\bibnamefont {Hu}}, \bibinfo {author} {\bibfnamefont {P.}~\bibnamefont {Jiang}}, \bibinfo {author} {\bibfnamefont {D.~H.}\ \bibnamefont {Kim}}, \bibinfo {author} {\bibfnamefont {G.~F.}\ \bibnamefont {Dionne}}, \bibinfo {author} {\bibfnamefont {L.~C.}\ \bibnamefont {Kimerling}},\ and\ \bibinfo {author} {\bibfnamefont {A.}~\bibnamefont {RossC.}},\ }\bibfield  {title} {\bibinfo {title} {On-chip optical isolation in monolithically integrated non-reciprocal optical resonators},\ }\href {http://dx.doi.org/10.1038/nphoton.2011.270} {\bibfield  {journal} {\bibinfo  {journal} {Nat Photon}\ }\textbf {\bibinfo {volume} {5}},\ \bibinfo {pages} {758} (\bibinfo {year} {2011})}\BibitemShut {NoStop}%
\bibitem [{\citenamefont {Kang}\ \emph {et~al.}(2011)\citenamefont {Kang}, \citenamefont {Butsch},\ and\ \citenamefont {Russell}}]{KangM.2011}%
  \BibitemOpen
  \bibfield  {author} {\bibinfo {author} {\bibfnamefont {M.~S.}\ \bibnamefont {Kang}}, \bibinfo {author} {\bibfnamefont {A.}~\bibnamefont {Butsch}},\ and\ \bibinfo {author} {\bibfnamefont {P.~S.~J.}\ \bibnamefont {Russell}},\ }\bibfield  {title} {\bibinfo {title} {Reconfigurable light-driven opto-acoustic isolators in photonic crystal fibre},\ }\href {http://dx.doi.org/10.1038/nphoton.2011.180} {\bibfield  {journal} {\bibinfo  {journal} {Nat Photon}\ }\textbf {\bibinfo {volume} {5}},\ \bibinfo {pages} {549} (\bibinfo {year} {2011})}\BibitemShut {NoStop}%
\bibitem [{\citenamefont {Hafezi}\ and\ \citenamefont {Rabl}(2012)}]{Hafezi2012a}%
  \BibitemOpen
  \bibfield  {author} {\bibinfo {author} {\bibfnamefont {M.}~\bibnamefont {Hafezi}}\ and\ \bibinfo {author} {\bibfnamefont {P.}~\bibnamefont {Rabl}},\ }\bibfield  {title} {\bibinfo {title} {Optomechanically induced non-reciprocity in microring resonators},\ }\bibfield  {booktitle} {\emph {\bibinfo {booktitle} {Optics Express}},\ }\href {http://www.opticsexpress.org/abstract.cfm?URI=oe-20-7-7672} {\bibfield  {journal} {\bibinfo  {journal} {Opt. Express}\ }\textbf {\bibinfo {volume} {20}},\ \bibinfo {pages} {7672} (\bibinfo {year} {2012})}\BibitemShut {NoStop}%
\bibitem [{\citenamefont {Metelmann}\ and\ \citenamefont {Clerk}(2015{\natexlab{a}})}]{Metelmann2015}%
  \BibitemOpen
  \bibfield  {author} {\bibinfo {author} {\bibfnamefont {A.}~\bibnamefont {Metelmann}}\ and\ \bibinfo {author} {\bibfnamefont {A.~A.}\ \bibnamefont {Clerk}},\ }\bibfield  {title} {\bibinfo {title} {Nonreciprocal photon transmission and amplification via reservoir engineering},\ }\href {https://doi.org/10.1103/PhysRevX.5.021025} {\bibfield  {journal} {\bibinfo  {journal} {Phys. Rev. X}\ }\textbf {\bibinfo {volume} {5}},\ \bibinfo {pages} {021025} (\bibinfo {year} {2015}{\natexlab{a}})}\BibitemShut {NoStop}%
\bibitem [{\citenamefont {Xu}\ and\ \citenamefont {Li}(2015)}]{Xu2015}%
  \BibitemOpen
  \bibfield  {author} {\bibinfo {author} {\bibfnamefont {X.-W.}\ \bibnamefont {Xu}}\ and\ \bibinfo {author} {\bibfnamefont {Y.}~\bibnamefont {Li}},\ }\bibfield  {title} {\bibinfo {title} {Optical nonreciprocity and optomechanical circulator in three-mode optomechanical systems},\ }\href {https://doi.org/10.1103/PhysRevA.91.053854} {\bibfield  {journal} {\bibinfo  {journal} {Phys. Rev. A}\ }\textbf {\bibinfo {volume} {91}},\ \bibinfo {pages} {053854} (\bibinfo {year} {2015})}\BibitemShut {NoStop}%
\bibitem [{\citenamefont {Shen}\ \emph {et~al.}(2016)\citenamefont {Shen}, \citenamefont {Zhang}, \citenamefont {Chen}, \citenamefont {Zou}, \citenamefont {Xiao}, \citenamefont {Zou}, \citenamefont {Sun}, \citenamefont {Guo},\ and\ \citenamefont {Dong}}]{Shen2016}%
  \BibitemOpen
  \bibfield  {author} {\bibinfo {author} {\bibfnamefont {Z.}~\bibnamefont {Shen}}, \bibinfo {author} {\bibfnamefont {Y.-L.}\ \bibnamefont {Zhang}}, \bibinfo {author} {\bibfnamefont {Y.}~\bibnamefont {Chen}}, \bibinfo {author} {\bibfnamefont {C.-L.}\ \bibnamefont {Zou}}, \bibinfo {author} {\bibfnamefont {Y.-F.}\ \bibnamefont {Xiao}}, \bibinfo {author} {\bibfnamefont {X.-B.}\ \bibnamefont {Zou}}, \bibinfo {author} {\bibfnamefont {F.-W.}\ \bibnamefont {Sun}}, \bibinfo {author} {\bibfnamefont {G.-C.}\ \bibnamefont {Guo}},\ and\ \bibinfo {author} {\bibfnamefont {C.-H.}\ \bibnamefont {Dong}},\ }\bibfield  {title} {\bibinfo {title} {Experimental realization of optomechanically induced non-reciprocity},\ }\href {http://dx.doi.org/10.1038/nphoton.2016.161} {\bibfield  {journal} {\bibinfo  {journal} {Nat Photon}\ }\textbf {\bibinfo {volume} {10}},\ \bibinfo {pages} {657} (\bibinfo {year} {2016})}\BibitemShut {NoStop}%
\bibitem [{\citenamefont {Ruesink}\ \emph {et~al.}(2016)\citenamefont {Ruesink}, \citenamefont {Miri}, \citenamefont {Alù},\ and\ \citenamefont {Verhagen}}]{Ruesink2016}%
  \BibitemOpen
  \bibfield  {author} {\bibinfo {author} {\bibfnamefont {F.}~\bibnamefont {Ruesink}}, \bibinfo {author} {\bibfnamefont {M.-A.}\ \bibnamefont {Miri}}, \bibinfo {author} {\bibfnamefont {A.}~\bibnamefont {Alù}},\ and\ \bibinfo {author} {\bibfnamefont {E.}~\bibnamefont {Verhagen}},\ }\bibfield  {title} {\bibinfo {title} {Nonreciprocity and magnetic-free isolation based on optomechanical interactions},\ }\href {http://dx.doi.org/10.1038/ncomms13662} {\bibfield  {journal} {\bibinfo  {journal} {Nature Communications}\ }\textbf {\bibinfo {volume} {7}},\ \bibinfo {pages} {13662} (\bibinfo {year} {2016})}\BibitemShut {NoStop}%
\bibitem [{\citenamefont {Fang}\ \emph {et~al.}(2017)\citenamefont {Fang}, \citenamefont {Luo}, \citenamefont {Metelmann}, \citenamefont {Matheny}, \citenamefont {Marquardt}, \citenamefont {Clerk},\ and\ \citenamefont {Painter}}]{Fang2017}%
  \BibitemOpen
  \bibfield  {author} {\bibinfo {author} {\bibfnamefont {K.}~\bibnamefont {Fang}}, \bibinfo {author} {\bibfnamefont {J.}~\bibnamefont {Luo}}, \bibinfo {author} {\bibfnamefont {A.}~\bibnamefont {Metelmann}}, \bibinfo {author} {\bibfnamefont {M.~H.}\ \bibnamefont {Matheny}}, \bibinfo {author} {\bibfnamefont {F.}~\bibnamefont {Marquardt}}, \bibinfo {author} {\bibfnamefont {A.~A.}\ \bibnamefont {Clerk}},\ and\ \bibinfo {author} {\bibfnamefont {O.}~\bibnamefont {Painter}},\ }\bibfield  {title} {\bibinfo {title} {Generalized non-reciprocity in an optomechanical circuit via synthetic magnetism and reservoir engineering},\ }\href {http://dx.doi.org/10.1038/nphys4009} {\bibfield  {journal} {\bibinfo  {journal} {Nat Phys}\ }\textbf {\bibinfo {volume} {13}},\ \bibinfo {pages} {465} (\bibinfo {year} {2017})}\BibitemShut {NoStop}%
\bibitem [{\citenamefont {Toth}\ \emph {et~al.}(2017)\citenamefont {Toth}, \citenamefont {Bernier}, \citenamefont {Nunnenkamp}, \citenamefont {Feofanov},\ and\ \citenamefont {Kippenberg}}]{Toth2017}%
  \BibitemOpen
  \bibfield  {author} {\bibinfo {author} {\bibfnamefont {L.~D.}\ \bibnamefont {Toth}}, \bibinfo {author} {\bibfnamefont {N.~R.}\ \bibnamefont {Bernier}}, \bibinfo {author} {\bibfnamefont {A.}~\bibnamefont {Nunnenkamp}}, \bibinfo {author} {\bibfnamefont {A.~K.}\ \bibnamefont {Feofanov}},\ and\ \bibinfo {author} {\bibfnamefont {T.~J.}\ \bibnamefont {Kippenberg}},\ }\bibfield  {title} {\bibinfo {title} {A dissipative quantum reservoir for microwave light using a mechanical oscillator},\ }\href {http://dx.doi.org/10.1038/nphys4121} {\bibfield  {journal} {\bibinfo  {journal} {Nat Phys}\ }\textbf {\bibinfo {volume} {advance online publication}},\  (\bibinfo {year} {2017})}\BibitemShut {NoStop}%
\bibitem [{\citenamefont {Barzanjeh}\ \emph {et~al.}(2017)\citenamefont {Barzanjeh}, \citenamefont {Wulf}, \citenamefont {Peruzzo}, \citenamefont {Kalaee}, \citenamefont {Dieterle}, \citenamefont {Painter},\ and\ \citenamefont {Fink}}]{Barzanjeh2017}%
  \BibitemOpen
  \bibfield  {author} {\bibinfo {author} {\bibfnamefont {S.}~\bibnamefont {Barzanjeh}}, \bibinfo {author} {\bibfnamefont {M.}~\bibnamefont {Wulf}}, \bibinfo {author} {\bibfnamefont {M.}~\bibnamefont {Peruzzo}}, \bibinfo {author} {\bibfnamefont {M.}~\bibnamefont {Kalaee}}, \bibinfo {author} {\bibfnamefont {P.~B.}\ \bibnamefont {Dieterle}}, \bibinfo {author} {\bibfnamefont {O.}~\bibnamefont {Painter}},\ and\ \bibinfo {author} {\bibfnamefont {J.~M.}\ \bibnamefont {Fink}},\ }\bibfield  {title} {\bibinfo {title} {Mechanical on-chip microwave circulator},\ }\href {https://doi.org/10.1038/s41467-017-01304-x} {\bibfield  {journal} {\bibinfo  {journal} {Nature Communications}\ }\textbf {\bibinfo {volume} {8}},\ \bibinfo {pages} {953} (\bibinfo {year} {2017})}\BibitemShut {NoStop}%
\bibitem [{\citenamefont {Barzanjeh}\ \emph {et~al.}(2018)\citenamefont {Barzanjeh}, \citenamefont {Aquilina},\ and\ \citenamefont {Xuereb}}]{PhysRevLett.120.060601}%
  \BibitemOpen
  \bibfield  {author} {\bibinfo {author} {\bibfnamefont {S.}~\bibnamefont {Barzanjeh}}, \bibinfo {author} {\bibfnamefont {M.}~\bibnamefont {Aquilina}},\ and\ \bibinfo {author} {\bibfnamefont {A.}~\bibnamefont {Xuereb}},\ }\bibfield  {title} {\bibinfo {title} {Manipulating the flow of thermal noise in quantum devices},\ }\href {https://doi.org/10.1103/PhysRevLett.120.060601} {\bibfield  {journal} {\bibinfo  {journal} {Phys. Rev. Lett.}\ }\textbf {\bibinfo {volume} {120}},\ \bibinfo {pages} {060601} (\bibinfo {year} {2018})}\BibitemShut {NoStop}%
\bibitem [{\citenamefont {Campaioli}\ \emph {et~al.}(2023)\citenamefont {Campaioli}, \citenamefont {Gherardini}, \citenamefont {Quach}, \citenamefont {Polini},\ and\ \citenamefont {Andolina}}]{Campaioli2023}%
  \BibitemOpen
  \bibfield  {author} {\bibinfo {author} {\bibfnamefont {F.}~\bibnamefont {Campaioli}}, \bibinfo {author} {\bibfnamefont {S.}~\bibnamefont {Gherardini}}, \bibinfo {author} {\bibfnamefont {J.~Q.}\ \bibnamefont {Quach}}, \bibinfo {author} {\bibfnamefont {M.}~\bibnamefont {Polini}},\ and\ \bibinfo {author} {\bibfnamefont {G.~M.}\ \bibnamefont {Andolina}},\ }\href@noop {} {\bibinfo {title} {Colloquium: Quantum batteries}} (\bibinfo {year} {2023}),\ \Eprint {https://arxiv.org/abs/2308.02277} {arXiv:2308.02277 [quant-ph]} \BibitemShut {NoStop}%
\bibitem [{\citenamefont {Alicki}\ and\ \citenamefont {Fannes}(2013)}]{PhysRevE.87.042123}%
  \BibitemOpen
  \bibfield  {author} {\bibinfo {author} {\bibfnamefont {R.}~\bibnamefont {Alicki}}\ and\ \bibinfo {author} {\bibfnamefont {M.}~\bibnamefont {Fannes}},\ }\bibfield  {title} {\bibinfo {title} {Entanglement boost for extractable work from ensembles of quantum batteries},\ }\href {https://doi.org/10.1103/PhysRevE.87.042123} {\bibfield  {journal} {\bibinfo  {journal} {Phys. Rev. E}\ }\textbf {\bibinfo {volume} {87}},\ \bibinfo {pages} {042123} (\bibinfo {year} {2013})}\BibitemShut {NoStop}%
\bibitem [{\citenamefont {Campaioli}\ \emph {et~al.}(2017)\citenamefont {Campaioli}, \citenamefont {Pollock}, \citenamefont {Binder}, \citenamefont {C\'eleri}, \citenamefont {Goold}, \citenamefont {Vinjanampathy},\ and\ \citenamefont {Modi}}]{PhysRevLett.118.150601}%
  \BibitemOpen
  \bibfield  {author} {\bibinfo {author} {\bibfnamefont {F.}~\bibnamefont {Campaioli}}, \bibinfo {author} {\bibfnamefont {F.~A.}\ \bibnamefont {Pollock}}, \bibinfo {author} {\bibfnamefont {F.~C.}\ \bibnamefont {Binder}}, \bibinfo {author} {\bibfnamefont {L.}~\bibnamefont {C\'eleri}}, \bibinfo {author} {\bibfnamefont {J.}~\bibnamefont {Goold}}, \bibinfo {author} {\bibfnamefont {S.}~\bibnamefont {Vinjanampathy}},\ and\ \bibinfo {author} {\bibfnamefont {K.}~\bibnamefont {Modi}},\ }\bibfield  {title} {\bibinfo {title} {Enhancing the charging power of quantum batteries},\ }\href {https://doi.org/10.1103/PhysRevLett.118.150601} {\bibfield  {journal} {\bibinfo  {journal} {Phys. Rev. Lett.}\ }\textbf {\bibinfo {volume} {118}},\ \bibinfo {pages} {150601} (\bibinfo {year} {2017})}\BibitemShut {NoStop}%
\bibitem [{\citenamefont {Andolina}\ \emph {et~al.}(2019{\natexlab{a}})\citenamefont {Andolina}, \citenamefont {Keck}, \citenamefont {Mari}, \citenamefont {Giovannetti},\ and\ \citenamefont {Polini}}]{PhysRevB.99.205437}%
  \BibitemOpen
  \bibfield  {author} {\bibinfo {author} {\bibfnamefont {G.~M.}\ \bibnamefont {Andolina}}, \bibinfo {author} {\bibfnamefont {M.}~\bibnamefont {Keck}}, \bibinfo {author} {\bibfnamefont {A.}~\bibnamefont {Mari}}, \bibinfo {author} {\bibfnamefont {V.}~\bibnamefont {Giovannetti}},\ and\ \bibinfo {author} {\bibfnamefont {M.}~\bibnamefont {Polini}},\ }\bibfield  {title} {\bibinfo {title} {Quantum versus classical many-body batteries},\ }\href {https://doi.org/10.1103/PhysRevB.99.205437} {\bibfield  {journal} {\bibinfo  {journal} {Phys. Rev. B}\ }\textbf {\bibinfo {volume} {99}},\ \bibinfo {pages} {205437} (\bibinfo {year} {2019}{\natexlab{a}})}\BibitemShut {NoStop}%
\bibitem [{\citenamefont {Binder}\ \emph {et~al.}(2015)\citenamefont {Binder}, \citenamefont {Vinjanampathy}, \citenamefont {Modi},\ and\ \citenamefont {Goold}}]{binder2015quantacell}%
  \BibitemOpen
  \bibfield  {author} {\bibinfo {author} {\bibfnamefont {F.~C.}\ \bibnamefont {Binder}}, \bibinfo {author} {\bibfnamefont {S.}~\bibnamefont {Vinjanampathy}}, \bibinfo {author} {\bibfnamefont {K.}~\bibnamefont {Modi}},\ and\ \bibinfo {author} {\bibfnamefont {J.}~\bibnamefont {Goold}},\ }\bibfield  {title} {\bibinfo {title} {Quantacell: powerful charging of quantum batteries},\ }\href {https://doi.org/10.1088/1367-2630/17/7/075015} {\bibfield  {journal} {\bibinfo  {journal} {New Journal of Physics}\ }\textbf {\bibinfo {volume} {17}},\ \bibinfo {pages} {075015} (\bibinfo {year} {2015})}\BibitemShut {NoStop}%
\bibitem [{\citenamefont {Farina}\ \emph {et~al.}(2019)\citenamefont {Farina}, \citenamefont {Andolina}, \citenamefont {Mari}, \citenamefont {Polini},\ and\ \citenamefont {Giovannetti}}]{farina2019charger}%
  \BibitemOpen
  \bibfield  {author} {\bibinfo {author} {\bibfnamefont {D.}~\bibnamefont {Farina}}, \bibinfo {author} {\bibfnamefont {G.~M.}\ \bibnamefont {Andolina}}, \bibinfo {author} {\bibfnamefont {A.}~\bibnamefont {Mari}}, \bibinfo {author} {\bibfnamefont {M.}~\bibnamefont {Polini}},\ and\ \bibinfo {author} {\bibfnamefont {V.}~\bibnamefont {Giovannetti}},\ }\bibfield  {title} {\bibinfo {title} {Charger-mediated energy transfer for quantum batteries: An open-system approach},\ }\href {https://doi.org/10.1103/PhysRevB.99.035421} {\bibfield  {journal} {\bibinfo  {journal} {Phys. Rev. B}\ }\textbf {\bibinfo {volume} {99}},\ \bibinfo {pages} {035421} (\bibinfo {year} {2019})}\BibitemShut {NoStop}%
\bibitem [{\citenamefont {Hovhannisyan}\ \emph {et~al.}(2013)\citenamefont {Hovhannisyan}, \citenamefont {Perarnau-Llobet}, \citenamefont {Huber},\ and\ \citenamefont {Ac\'{\i}n}}]{PhysRevLett.111.240401}%
  \BibitemOpen
  \bibfield  {author} {\bibinfo {author} {\bibfnamefont {K.~V.}\ \bibnamefont {Hovhannisyan}}, \bibinfo {author} {\bibfnamefont {M.}~\bibnamefont {Perarnau-Llobet}}, \bibinfo {author} {\bibfnamefont {M.}~\bibnamefont {Huber}},\ and\ \bibinfo {author} {\bibfnamefont {A.}~\bibnamefont {Ac\'{\i}n}},\ }\bibfield  {title} {\bibinfo {title} {Entanglement generation is not necessary for optimal work extraction},\ }\href {https://doi.org/10.1103/PhysRevLett.111.240401} {\bibfield  {journal} {\bibinfo  {journal} {Phys. Rev. Lett.}\ }\textbf {\bibinfo {volume} {111}},\ \bibinfo {pages} {240401} (\bibinfo {year} {2013})}\BibitemShut {NoStop}%
\bibitem [{\citenamefont {Ferraro}\ \emph {et~al.}(2018)\citenamefont {Ferraro}, \citenamefont {Campisi}, \citenamefont {Andolina}, \citenamefont {Pellegrini},\ and\ \citenamefont {Polini}}]{PhysRevLett.120.117702}%
  \BibitemOpen
  \bibfield  {author} {\bibinfo {author} {\bibfnamefont {D.}~\bibnamefont {Ferraro}}, \bibinfo {author} {\bibfnamefont {M.}~\bibnamefont {Campisi}}, \bibinfo {author} {\bibfnamefont {G.~M.}\ \bibnamefont {Andolina}}, \bibinfo {author} {\bibfnamefont {V.}~\bibnamefont {Pellegrini}},\ and\ \bibinfo {author} {\bibfnamefont {M.}~\bibnamefont {Polini}},\ }\bibfield  {title} {\bibinfo {title} {High-power collective charging of a solid-state quantum battery},\ }\href {https://doi.org/10.1103/PhysRevLett.120.117702} {\bibfield  {journal} {\bibinfo  {journal} {Phys. Rev. Lett.}\ }\textbf {\bibinfo {volume} {120}},\ \bibinfo {pages} {117702} (\bibinfo {year} {2018})}\BibitemShut {NoStop}%
\bibitem [{\citenamefont {Quach}\ \emph {et~al.}(2022{\natexlab{a}})\citenamefont {Quach}, \citenamefont {McGhee}, \citenamefont {Ganzer}, \citenamefont {Rouse}, \citenamefont {Lovett}, \citenamefont {Gauger}, \citenamefont {Keeling}, \citenamefont {Cerullo}, \citenamefont {Lidzey},\ and\ \citenamefont {Virgili}}]{eabk3160}%
  \BibitemOpen
  \bibfield  {author} {\bibinfo {author} {\bibfnamefont {J.~Q.}\ \bibnamefont {Quach}}, \bibinfo {author} {\bibfnamefont {K.~E.}\ \bibnamefont {McGhee}}, \bibinfo {author} {\bibfnamefont {L.}~\bibnamefont {Ganzer}}, \bibinfo {author} {\bibfnamefont {D.~M.}\ \bibnamefont {Rouse}}, \bibinfo {author} {\bibfnamefont {B.~W.}\ \bibnamefont {Lovett}}, \bibinfo {author} {\bibfnamefont {E.~M.}\ \bibnamefont {Gauger}}, \bibinfo {author} {\bibfnamefont {J.}~\bibnamefont {Keeling}}, \bibinfo {author} {\bibfnamefont {G.}~\bibnamefont {Cerullo}}, \bibinfo {author} {\bibfnamefont {D.~G.}\ \bibnamefont {Lidzey}},\ and\ \bibinfo {author} {\bibfnamefont {T.}~\bibnamefont {Virgili}},\ }\bibfield  {title} {\bibinfo {title} {Superabsorption in an organic microcavity: Toward a quantum battery},\ }\href {https://doi.org/10.1126/sciadv.abk3160} {\bibfield  {journal} {\bibinfo  {journal} {Science Advances}\ }\textbf {\bibinfo {volume} {8}},\ \bibinfo {pages} {eabk3160} (\bibinfo {year} {2022}{\natexlab{a}})},\ \Eprint
  {https://arxiv.org/abs/https://www.science.org/doi/pdf/10.1126/sciadv.abk3160} {https://www.science.org/doi/pdf/10.1126/sciadv.abk3160} \BibitemShut {NoStop}%
\bibitem [{\citenamefont {Barra}(2019)}]{Barra}%
  \BibitemOpen
  \bibfield  {author} {\bibinfo {author} {\bibfnamefont {F.}~\bibnamefont {Barra}},\ }\bibfield  {title} {\bibinfo {title} {Dissipative charging of a quantum battery},\ }\href {https://doi.org/10.1103/PhysRevLett.122.210601} {\bibfield  {journal} {\bibinfo  {journal} {Phys. Rev. Lett.}\ }\textbf {\bibinfo {volume} {122}},\ \bibinfo {pages} {210601} (\bibinfo {year} {2019})}\BibitemShut {NoStop}%
\bibitem [{\citenamefont {Bhattacharjee}\ and\ \citenamefont {Dutta}(2021)}]{Bhattacharjee2021}%
  \BibitemOpen
  \bibfield  {author} {\bibinfo {author} {\bibfnamefont {S.}~\bibnamefont {Bhattacharjee}}\ and\ \bibinfo {author} {\bibfnamefont {A.}~\bibnamefont {Dutta}},\ }\bibfield  {title} {\bibinfo {title} {Quantum thermal machines and batteries},\ }\href {https://doi.org/10.1140/epjb/s10051-021-00235-3} {\bibfield  {journal} {\bibinfo  {journal} {The European Physical Journal B}\ }\textbf {\bibinfo {volume} {94}},\ \bibinfo {pages} {239} (\bibinfo {year} {2021})}\BibitemShut {NoStop}%
\bibitem [{\citenamefont {Quach}\ and\ \citenamefont {Munro}(2020)}]{PhysRevApplied.14.024092}%
  \BibitemOpen
  \bibfield  {author} {\bibinfo {author} {\bibfnamefont {J.~Q.}\ \bibnamefont {Quach}}\ and\ \bibinfo {author} {\bibfnamefont {W.~J.}\ \bibnamefont {Munro}},\ }\bibfield  {title} {\bibinfo {title} {Using dark states to charge and stabilize open quantum batteries},\ }\href {https://doi.org/10.1103/PhysRevApplied.14.024092} {\bibfield  {journal} {\bibinfo  {journal} {Phys. Rev. Appl.}\ }\textbf {\bibinfo {volume} {14}},\ \bibinfo {pages} {024092} (\bibinfo {year} {2020})}\BibitemShut {NoStop}%
\bibitem [{\citenamefont {Andolina}\ \emph {et~al.}(2019{\natexlab{b}})\citenamefont {Andolina}, \citenamefont {Keck}, \citenamefont {Mari}, \citenamefont {Campisi}, \citenamefont {Giovannetti},\ and\ \citenamefont {Polini}}]{PhysRevLett.122.047702}%
  \BibitemOpen
  \bibfield  {author} {\bibinfo {author} {\bibfnamefont {G.~M.}\ \bibnamefont {Andolina}}, \bibinfo {author} {\bibfnamefont {M.}~\bibnamefont {Keck}}, \bibinfo {author} {\bibfnamefont {A.}~\bibnamefont {Mari}}, \bibinfo {author} {\bibfnamefont {M.}~\bibnamefont {Campisi}}, \bibinfo {author} {\bibfnamefont {V.}~\bibnamefont {Giovannetti}},\ and\ \bibinfo {author} {\bibfnamefont {M.}~\bibnamefont {Polini}},\ }\bibfield  {title} {\bibinfo {title} {Extractable work, the role of correlations, and asymptotic freedom in quantum batteries},\ }\href {https://doi.org/10.1103/PhysRevLett.122.047702} {\bibfield  {journal} {\bibinfo  {journal} {Phys. Rev. Lett.}\ }\textbf {\bibinfo {volume} {122}},\ \bibinfo {pages} {047702} (\bibinfo {year} {2019}{\natexlab{b}})}\BibitemShut {NoStop}%
\bibitem [{\citenamefont {Salvia}\ \emph {et~al.}(2023)\citenamefont {Salvia}, \citenamefont {Perarnau-Llobet}, \citenamefont {Haack}, \citenamefont {Brunner},\ and\ \citenamefont {Nimmrichter}}]{PhysRevResearch.5.013155}%
  \BibitemOpen
  \bibfield  {author} {\bibinfo {author} {\bibfnamefont {R.}~\bibnamefont {Salvia}}, \bibinfo {author} {\bibfnamefont {M.}~\bibnamefont {Perarnau-Llobet}}, \bibinfo {author} {\bibfnamefont {G.}~\bibnamefont {Haack}}, \bibinfo {author} {\bibfnamefont {N.}~\bibnamefont {Brunner}},\ and\ \bibinfo {author} {\bibfnamefont {S.}~\bibnamefont {Nimmrichter}},\ }\bibfield  {title} {\bibinfo {title} {Quantum advantage in charging cavity and spin batteries by repeated interactions},\ }\href {https://doi.org/10.1103/PhysRevResearch.5.013155} {\bibfield  {journal} {\bibinfo  {journal} {Phys. Rev. Res.}\ }\textbf {\bibinfo {volume} {5}},\ \bibinfo {pages} {013155} (\bibinfo {year} {2023})}\BibitemShut {NoStop}%
\bibitem [{\citenamefont {Rodriguez}\ \emph {et~al.}(2022)\citenamefont {Rodriguez}, \citenamefont {Ahmadi}, \citenamefont {Suarez}, \citenamefont {Mazurek}, \citenamefont {Barzanjeh},\ and\ \citenamefont {Horodecki}}]{rodriguez2022optimal}%
  \BibitemOpen
  \bibfield  {author} {\bibinfo {author} {\bibfnamefont {R.}~\bibnamefont {Rodriguez}}, \bibinfo {author} {\bibfnamefont {B.}~\bibnamefont {Ahmadi}}, \bibinfo {author} {\bibfnamefont {G.}~\bibnamefont {Suarez}}, \bibinfo {author} {\bibfnamefont {P.}~\bibnamefont {Mazurek}}, \bibinfo {author} {\bibfnamefont {S.}~\bibnamefont {Barzanjeh}},\ and\ \bibinfo {author} {\bibfnamefont {P.}~\bibnamefont {Horodecki}},\ }\bibfield  {title} {\bibinfo {title} {Optimal quantum control of charging quantum batteries},\ }\bibfield  {journal} {\bibinfo  {journal} {arXiv preprint arXiv:2207.00094}\ }\href {https://doi.org/arXiv.2207.00094} {arXiv.2207.00094} (\bibinfo {year} {2022})\BibitemShut {NoStop}%
\bibitem [{\citenamefont {Mazzoncini}\ \emph {et~al.}(2023)\citenamefont {Mazzoncini}, \citenamefont {Cavina}, \citenamefont {Andolina}, \citenamefont {Erdman},\ and\ \citenamefont {Giovannetti}}]{PhysRevA.107.032218}%
  \BibitemOpen
  \bibfield  {author} {\bibinfo {author} {\bibfnamefont {F.}~\bibnamefont {Mazzoncini}}, \bibinfo {author} {\bibfnamefont {V.}~\bibnamefont {Cavina}}, \bibinfo {author} {\bibfnamefont {G.~M.}\ \bibnamefont {Andolina}}, \bibinfo {author} {\bibfnamefont {P.~A.}\ \bibnamefont {Erdman}},\ and\ \bibinfo {author} {\bibfnamefont {V.}~\bibnamefont {Giovannetti}},\ }\bibfield  {title} {\bibinfo {title} {Optimal control methods for quantum batteries},\ }\href {https://doi.org/10.1103/PhysRevA.107.032218} {\bibfield  {journal} {\bibinfo  {journal} {Phys. Rev. A}\ }\textbf {\bibinfo {volume} {107}},\ \bibinfo {pages} {032218} (\bibinfo {year} {2023})}\BibitemShut {NoStop}%
\bibitem [{\citenamefont {Rodr\'{\i}guez}\ \emph {et~al.}(2023)\citenamefont {Rodr\'{\i}guez}, \citenamefont {Ahmadi}, \citenamefont {Mazurek}, \citenamefont {Barzanjeh}, \citenamefont {Alicki},\ and\ \citenamefont {Horodecki}}]{PhysRevA.107.042419}%
  \BibitemOpen
  \bibfield  {author} {\bibinfo {author} {\bibfnamefont {R.~R.}\ \bibnamefont {Rodr\'{\i}guez}}, \bibinfo {author} {\bibfnamefont {B.}~\bibnamefont {Ahmadi}}, \bibinfo {author} {\bibfnamefont {P.}~\bibnamefont {Mazurek}}, \bibinfo {author} {\bibfnamefont {S.}~\bibnamefont {Barzanjeh}}, \bibinfo {author} {\bibfnamefont {R.}~\bibnamefont {Alicki}},\ and\ \bibinfo {author} {\bibfnamefont {P.}~\bibnamefont {Horodecki}},\ }\bibfield  {title} {\bibinfo {title} {Catalysis in charging quantum batteries},\ }\href {https://doi.org/10.1103/PhysRevA.107.042419} {\bibfield  {journal} {\bibinfo  {journal} {Phys. Rev. A}\ }\textbf {\bibinfo {volume} {107}},\ \bibinfo {pages} {042419} (\bibinfo {year} {2023})}\BibitemShut {NoStop}%
\bibitem [{\citenamefont {Quach}\ \emph {et~al.}(2022{\natexlab{b}})\citenamefont {Quach}, \citenamefont {McGhee}, \citenamefont {Ganzer}, \citenamefont {Rouse}, \citenamefont {Lovett}, \citenamefont {Gauger}, \citenamefont {Keeling}, \citenamefont {Cerullo}, \citenamefont {Lidzey},\ and\ \citenamefont {Virgili}}]{Quach}%
  \BibitemOpen
  \bibfield  {author} {\bibinfo {author} {\bibfnamefont {J.~Q.}\ \bibnamefont {Quach}}, \bibinfo {author} {\bibfnamefont {K.~E.}\ \bibnamefont {McGhee}}, \bibinfo {author} {\bibfnamefont {L.}~\bibnamefont {Ganzer}}, \bibinfo {author} {\bibfnamefont {D.~M.}\ \bibnamefont {Rouse}}, \bibinfo {author} {\bibfnamefont {B.~W.}\ \bibnamefont {Lovett}}, \bibinfo {author} {\bibfnamefont {E.~M.}\ \bibnamefont {Gauger}}, \bibinfo {author} {\bibfnamefont {J.}~\bibnamefont {Keeling}}, \bibinfo {author} {\bibfnamefont {G.}~\bibnamefont {Cerullo}}, \bibinfo {author} {\bibfnamefont {D.~G.}\ \bibnamefont {Lidzey}},\ and\ \bibinfo {author} {\bibfnamefont {T.}~\bibnamefont {Virgili}},\ }\bibfield  {title} {\bibinfo {title} {Superabsorption in an organic microcavity: Toward a quantum battery},\ }\href {https://doi.org/10.1126/sciadv.abk3160} {\bibfield  {journal} {\bibinfo  {journal} {Science Advances}\ }\textbf {\bibinfo {volume} {8}},\ \bibinfo {pages} {eabk3160} (\bibinfo {year} {2022}{\natexlab{b}})}\BibitemShut {NoStop}%
\bibitem [{\citenamefont {Hu}\ \emph {et~al.}(2022)\citenamefont {Hu}, \citenamefont {Qiu}, \citenamefont {Souza}, \citenamefont {Yuan}, \citenamefont {Zhou}, \citenamefont {Zhang}, \citenamefont {Chu}, \citenamefont {Pan}, \citenamefont {Hu}, \citenamefont {Li}, \citenamefont {Xu}, \citenamefont {Zhong}, \citenamefont {Liu}, \citenamefont {Yan}, \citenamefont {Tan}, \citenamefont {Bachelard}, \citenamefont {Villas-Boas}, \citenamefont {Santos},\ and\ \citenamefont {Yu}}]{Hu_2022}%
  \BibitemOpen
  \bibfield  {author} {\bibinfo {author} {\bibfnamefont {C.-K.}\ \bibnamefont {Hu}}, \bibinfo {author} {\bibfnamefont {J.}~\bibnamefont {Qiu}}, \bibinfo {author} {\bibfnamefont {P.~J.~P.}\ \bibnamefont {Souza}}, \bibinfo {author} {\bibfnamefont {J.}~\bibnamefont {Yuan}}, \bibinfo {author} {\bibfnamefont {Y.}~\bibnamefont {Zhou}}, \bibinfo {author} {\bibfnamefont {L.}~\bibnamefont {Zhang}}, \bibinfo {author} {\bibfnamefont {J.}~\bibnamefont {Chu}}, \bibinfo {author} {\bibfnamefont {X.}~\bibnamefont {Pan}}, \bibinfo {author} {\bibfnamefont {L.}~\bibnamefont {Hu}}, \bibinfo {author} {\bibfnamefont {J.}~\bibnamefont {Li}}, \bibinfo {author} {\bibfnamefont {Y.}~\bibnamefont {Xu}}, \bibinfo {author} {\bibfnamefont {Y.}~\bibnamefont {Zhong}}, \bibinfo {author} {\bibfnamefont {S.}~\bibnamefont {Liu}}, \bibinfo {author} {\bibfnamefont {F.}~\bibnamefont {Yan}}, \bibinfo {author} {\bibfnamefont {D.}~\bibnamefont {Tan}}, \bibinfo {author} {\bibfnamefont {R.}~\bibnamefont {Bachelard}}, \bibinfo {author} {\bibfnamefont
  {C.~J.}\ \bibnamefont {Villas-Boas}}, \bibinfo {author} {\bibfnamefont {A.~C.}\ \bibnamefont {Santos}},\ and\ \bibinfo {author} {\bibfnamefont {D.}~\bibnamefont {Yu}},\ }\bibfield  {title} {\bibinfo {title} {Optimal charging of a superconducting quantum battery},\ }\href {https://doi.org/10.1088/2058-9565/ac8444} {\bibfield  {journal} {\bibinfo  {journal} {Quantum Science and Technology}\ }\textbf {\bibinfo {volume} {7}},\ \bibinfo {pages} {045018} (\bibinfo {year} {2022})}\BibitemShut {NoStop}%
\bibitem [{\citenamefont {Joshi}\ and\ \citenamefont {Mahesh}(2022)}]{PhysRevA.106.042601}%
  \BibitemOpen
  \bibfield  {author} {\bibinfo {author} {\bibfnamefont {J.}~\bibnamefont {Joshi}}\ and\ \bibinfo {author} {\bibfnamefont {T.~S.}\ \bibnamefont {Mahesh}},\ }\bibfield  {title} {\bibinfo {title} {Experimental investigation of a quantum battery using star-topology nmr spin systems},\ }\href {https://doi.org/10.1103/PhysRevA.106.042601} {\bibfield  {journal} {\bibinfo  {journal} {Phys. Rev. A}\ }\textbf {\bibinfo {volume} {106}},\ \bibinfo {pages} {042601} (\bibinfo {year} {2022})}\BibitemShut {NoStop}%
\bibitem [{\citenamefont {Quach}\ \emph {et~al.}(2023)\citenamefont {Quach}, \citenamefont {Cerullo},\ and\ \citenamefont {Virgili}}]{QUACH20232195}%
  \BibitemOpen
  \bibfield  {author} {\bibinfo {author} {\bibfnamefont {J.}~\bibnamefont {Quach}}, \bibinfo {author} {\bibfnamefont {G.}~\bibnamefont {Cerullo}},\ and\ \bibinfo {author} {\bibfnamefont {T.}~\bibnamefont {Virgili}},\ }\bibfield  {title} {\bibinfo {title} {Quantum batteries: The future of energy storage?},\ }\href {https://doi.org/https://doi.org/10.1016/j.joule.2023.09.003} {\bibfield  {journal} {\bibinfo  {journal} {Joule}\ }\textbf {\bibinfo {volume} {7}},\ \bibinfo {pages} {2195} (\bibinfo {year} {2023})}\BibitemShut {NoStop}%
\bibitem [{\citenamefont {Poyatos}\ \emph {et~al.}(1996)\citenamefont {Poyatos}, \citenamefont {Cirac},\ and\ \citenamefont {Zoller}}]{PhysRevLett.77.4728}%
  \BibitemOpen
  \bibfield  {author} {\bibinfo {author} {\bibfnamefont {J.~F.}\ \bibnamefont {Poyatos}}, \bibinfo {author} {\bibfnamefont {J.~I.}\ \bibnamefont {Cirac}},\ and\ \bibinfo {author} {\bibfnamefont {P.}~\bibnamefont {Zoller}},\ }\bibfield  {title} {\bibinfo {title} {Quantum reservoir engineering with laser cooled trapped ions},\ }\href {https://doi.org/10.1103/PhysRevLett.77.4728} {\bibfield  {journal} {\bibinfo  {journal} {Phys. Rev. Lett.}\ }\textbf {\bibinfo {volume} {77}},\ \bibinfo {pages} {4728} (\bibinfo {year} {1996})}\BibitemShut {NoStop}%
\bibitem [{\citenamefont {Metelmann}\ and\ \citenamefont {Clerk}(2014)}]{PhysRevLett.112.133904}%
  \BibitemOpen
  \bibfield  {author} {\bibinfo {author} {\bibfnamefont {A.}~\bibnamefont {Metelmann}}\ and\ \bibinfo {author} {\bibfnamefont {A.~A.}\ \bibnamefont {Clerk}},\ }\bibfield  {title} {\bibinfo {title} {Quantum-limited amplification via reservoir engineering},\ }\href {https://doi.org/10.1103/PhysRevLett.112.133904} {\bibfield  {journal} {\bibinfo  {journal} {Phys. Rev. Lett.}\ }\textbf {\bibinfo {volume} {112}},\ \bibinfo {pages} {133904} (\bibinfo {year} {2014})}\BibitemShut {NoStop}%
\bibitem [{\citenamefont {Metelmann}\ and\ \citenamefont {Clerk}(2015{\natexlab{b}})}]{PhysRevX.5.021025}%
  \BibitemOpen
  \bibfield  {author} {\bibinfo {author} {\bibfnamefont {A.}~\bibnamefont {Metelmann}}\ and\ \bibinfo {author} {\bibfnamefont {A.~A.}\ \bibnamefont {Clerk}},\ }\bibfield  {title} {\bibinfo {title} {Nonreciprocal photon transmission and amplification via reservoir engineering},\ }\href {https://doi.org/10.1103/PhysRevX.5.021025} {\bibfield  {journal} {\bibinfo  {journal} {Phys. Rev. X}\ }\textbf {\bibinfo {volume} {5}},\ \bibinfo {pages} {021025} (\bibinfo {year} {2015}{\natexlab{b}})}\BibitemShut {NoStop}%
\bibitem [{\citenamefont {Kerckhoff}\ \emph {et~al.}(2015{\natexlab{b}})\citenamefont {Kerckhoff}, \citenamefont {Lalumi\`ere}, \citenamefont {Chapman}, \citenamefont {Blais},\ and\ \citenamefont {Lehnert}}]{PhysRevApplied.4.034002}%
  \BibitemOpen
  \bibfield  {author} {\bibinfo {author} {\bibfnamefont {J.}~\bibnamefont {Kerckhoff}}, \bibinfo {author} {\bibfnamefont {K.}~\bibnamefont {Lalumi\`ere}}, \bibinfo {author} {\bibfnamefont {B.~J.}\ \bibnamefont {Chapman}}, \bibinfo {author} {\bibfnamefont {A.}~\bibnamefont {Blais}},\ and\ \bibinfo {author} {\bibfnamefont {K.~W.}\ \bibnamefont {Lehnert}},\ }\bibfield  {title} {\bibinfo {title} {On-chip superconducting microwave circulator from synthetic rotation},\ }\href {https://doi.org/10.1103/PhysRevApplied.4.034002} {\bibfield  {journal} {\bibinfo  {journal} {Phys. Rev. Appl.}\ }\textbf {\bibinfo {volume} {4}},\ \bibinfo {pages} {034002} (\bibinfo {year} {2015}{\natexlab{b}})}\BibitemShut {NoStop}%
\bibitem [{\citenamefont {Chapman}\ \emph {et~al.}(2017{\natexlab{b}})\citenamefont {Chapman}, \citenamefont {Rosenthal}, \citenamefont {Kerckhoff}, \citenamefont {Moores}, \citenamefont {Vale}, \citenamefont {Mates}, \citenamefont {Hilton}, \citenamefont {Lalumi\`ere}, \citenamefont {Blais},\ and\ \citenamefont {Lehnert}}]{PhysRevX.7.041043}%
  \BibitemOpen
  \bibfield  {author} {\bibinfo {author} {\bibfnamefont {B.~J.}\ \bibnamefont {Chapman}}, \bibinfo {author} {\bibfnamefont {E.~I.}\ \bibnamefont {Rosenthal}}, \bibinfo {author} {\bibfnamefont {J.}~\bibnamefont {Kerckhoff}}, \bibinfo {author} {\bibfnamefont {B.~A.}\ \bibnamefont {Moores}}, \bibinfo {author} {\bibfnamefont {L.~R.}\ \bibnamefont {Vale}}, \bibinfo {author} {\bibfnamefont {J.~A.~B.}\ \bibnamefont {Mates}}, \bibinfo {author} {\bibfnamefont {G.~C.}\ \bibnamefont {Hilton}}, \bibinfo {author} {\bibfnamefont {K.}~\bibnamefont {Lalumi\`ere}}, \bibinfo {author} {\bibfnamefont {A.}~\bibnamefont {Blais}},\ and\ \bibinfo {author} {\bibfnamefont {K.~W.}\ \bibnamefont {Lehnert}},\ }\bibfield  {title} {\bibinfo {title} {Widely tunable on-chip microwave circulator for superconducting quantum circuits},\ }\href {https://doi.org/10.1103/PhysRevX.7.041043} {\bibfield  {journal} {\bibinfo  {journal} {Phys. Rev. X}\ }\textbf {\bibinfo {volume} {7}},\ \bibinfo {pages} {041043} (\bibinfo {year}
  {2017}{\natexlab{b}})}\BibitemShut {NoStop}%
\bibitem [{\citenamefont {Wang}\ \emph {et~al.}(2019)\citenamefont {Wang}, \citenamefont {Rao}, \citenamefont {Yang}, \citenamefont {Xu}, \citenamefont {Gui}, \citenamefont {Yao}, \citenamefont {You},\ and\ \citenamefont {Hu}}]{PhysRevLett.123.127202}%
  \BibitemOpen
  \bibfield  {author} {\bibinfo {author} {\bibfnamefont {Y.-P.}\ \bibnamefont {Wang}}, \bibinfo {author} {\bibfnamefont {J.~W.}\ \bibnamefont {Rao}}, \bibinfo {author} {\bibfnamefont {Y.}~\bibnamefont {Yang}}, \bibinfo {author} {\bibfnamefont {P.-C.}\ \bibnamefont {Xu}}, \bibinfo {author} {\bibfnamefont {Y.~S.}\ \bibnamefont {Gui}}, \bibinfo {author} {\bibfnamefont {B.~M.}\ \bibnamefont {Yao}}, \bibinfo {author} {\bibfnamefont {J.~Q.}\ \bibnamefont {You}},\ and\ \bibinfo {author} {\bibfnamefont {C.-M.}\ \bibnamefont {Hu}},\ }\bibfield  {title} {\bibinfo {title} {Nonreciprocity and unidirectional invisibility in cavity magnonics},\ }\href {https://doi.org/10.1103/PhysRevLett.123.127202} {\bibfield  {journal} {\bibinfo  {journal} {Phys. Rev. Lett.}\ }\textbf {\bibinfo {volume} {123}},\ \bibinfo {pages} {127202} (\bibinfo {year} {2019})}\BibitemShut {NoStop}%
\bibitem [{\citenamefont {Kim}\ \emph {et~al.}(2023)\citenamefont {Kim}, \citenamefont {Tabesh}, \citenamefont {Zegray}, \citenamefont {Barzanjeh},\ and\ \citenamefont {Hu}}]{2303.04358}%
  \BibitemOpen
  \bibfield  {author} {\bibinfo {author} {\bibfnamefont {M.}~\bibnamefont {Kim}}, \bibinfo {author} {\bibfnamefont {A.}~\bibnamefont {Tabesh}}, \bibinfo {author} {\bibfnamefont {T.}~\bibnamefont {Zegray}}, \bibinfo {author} {\bibfnamefont {S.}~\bibnamefont {Barzanjeh}},\ and\ \bibinfo {author} {\bibfnamefont {C.-M.}\ \bibnamefont {Hu}},\ }\href@noop {} {\bibinfo {title} {Nonreciprocity in cavity magnonics at milikelvin temperature}} (\bibinfo {year} {2023}),\ \Eprint {https://arxiv.org/abs/arXiv:2303.04358} {arXiv:2303.04358} \BibitemShut {NoStop}%
\bibitem [{\citenamefont {Andolina}\ \emph {et~al.}(2018)\citenamefont {Andolina}, \citenamefont {Farina}, \citenamefont {Mari}, \citenamefont {Pellegrini}, \citenamefont {Giovannetti},\ and\ \citenamefont {Polini}}]{Andolina2018}%
  \BibitemOpen
  \bibfield  {author} {\bibinfo {author} {\bibfnamefont {G.~M.}\ \bibnamefont {Andolina}}, \bibinfo {author} {\bibfnamefont {D.}~\bibnamefont {Farina}}, \bibinfo {author} {\bibfnamefont {A.}~\bibnamefont {Mari}}, \bibinfo {author} {\bibfnamefont {V.}~\bibnamefont {Pellegrini}}, \bibinfo {author} {\bibfnamefont {V.}~\bibnamefont {Giovannetti}},\ and\ \bibinfo {author} {\bibfnamefont {M.}~\bibnamefont {Polini}},\ }\bibfield  {title} {\bibinfo {title} {Charger-mediated energy transfer in exactly solvable models for quantum batteries},\ }\href {https://doi.org/10.1103/PhysRevB.98.205423} {\bibfield  {journal} {\bibinfo  {journal} {Phys. Rev. B}\ }\textbf {\bibinfo {volume} {98}},\ \bibinfo {pages} {205423} (\bibinfo {year} {2018})}\BibitemShut {NoStop}%
\bibitem [{\citenamefont {Kossakowski}(1972)}]{kossakowski1972quantum}%
  \BibitemOpen
  \bibfield  {author} {\bibinfo {author} {\bibfnamefont {A.}~\bibnamefont {Kossakowski}},\ }\bibfield  {title} {\bibinfo {title} {On quantum statistical mechanics of non-hamiltonian systems},\ }\href {https://doi.org/https://doi.org/10.1016/0034-4877(72)90010-9} {\bibfield  {journal} {\bibinfo  {journal} {Reports on Mathematical Physics}\ }\textbf {\bibinfo {volume} {3}},\ \bibinfo {pages} {247} (\bibinfo {year} {1972})}\BibitemShut {NoStop}%
\bibitem [{\citenamefont {Chiribella}\ \emph {et~al.}(2021)\citenamefont {Chiribella}, \citenamefont {Yang},\ and\ \citenamefont {Renner}}]{PhysRevX.11.021014}%
  \BibitemOpen
  \bibfield  {author} {\bibinfo {author} {\bibfnamefont {G.}~\bibnamefont {Chiribella}}, \bibinfo {author} {\bibfnamefont {Y.}~\bibnamefont {Yang}},\ and\ \bibinfo {author} {\bibfnamefont {R.}~\bibnamefont {Renner}},\ }\bibfield  {title} {\bibinfo {title} {Fundamental energy requirement of reversible quantum operations},\ }\href {https://doi.org/10.1103/PhysRevX.11.021014} {\bibfield  {journal} {\bibinfo  {journal} {Phys. Rev. X}\ }\textbf {\bibinfo {volume} {11}},\ \bibinfo {pages} {021014} (\bibinfo {year} {2021})}\BibitemShut {NoStop}%
\end{thebibliography}%
\clearpage

\onecolumngrid
\appendix
\begin{center}
\textbf{\large Appendix}
\end{center}

\section{Calculation of stationary energy in nonreciprocal regime}\label{SM1}

The stored energy in the battery in the nonreciprocal regime is computed by solving Eqs. (\ref{Ava})--(\ref{Avadb}) of the main text, resulting in
\begin{align*}\nonumber\label{AllTimeEnergyNonReciprocal2}
     E_{B}^{\text{nr}}(t):=\omega\langle b^{\dagger}b\rangle &_{nr}=\frac{16\omega\Gamma^2F^2}{\Phi_a \Phi_b} + \frac{64\delta^2\omega\Gamma^2F^2 }{\Phi_a \Phi_b (\Lambda_a-\Lambda_b)^2}\Big(  e^{-\frac{1}{2}\Lambda_a t} - e^{-\frac{1}{2}\Lambda_b t}\Big)^2 +\frac{16\omega\Gamma^2F^2 }{\Phi_a \Phi_b (\Lambda_a-\Lambda_b)^2}\Big( \Lambda_{a}e^{-\frac{1}{2}\Lambda_b t}-\Lambda_{b} e^{-\frac{1}{2}\Lambda_a t}\Big)^2\\
     &- \cos{\delta t}\frac{32\omega\Gamma^2F^2 }{\Phi_a \Phi_b (\Lambda_a-\Lambda_b)}\Big(  \Lambda_{a}e^{-\frac{1}{2}\Lambda_b t}-\Lambda_{b} e^{-\frac{1}{2}\Lambda_a t}\Big) + \delta\sin{\delta t}\frac{64\omega\Gamma^2F^2 }{\Phi_a \Phi_b (\Lambda_a - \Lambda_b)}\Big(  e^{-\frac{1}{2}\Lambda_a t}-e^{-\frac{1}{2}\Lambda_b t}\Big)
     ,\numberthis
\end{align*}
where $\delta=\omega_{L}-\omega$ is the detunning, $\Lambda_{a(b)}=\Gamma_{a(b)}+\kappa_{a(b)}$ and $\Phi_{a(b)}=\Lambda_{a(b)}^2+4\delta^2$. In the stationary limit, we always have 
\begin{equation}\label{StationaryEnergyBNonReciprocal}
\lim_{t\rightarrow\infty}E_B^{\text{nr}}(\delta=0)>\lim_{t\rightarrow\infty}E_B^{\text{nr}}(\delta\neq 0),    
\end{equation}
which justifies the resonant setting $\delta=0$ we apply in Eq. (\textcolor{blue}{5}) of the main text.

\section{Opimization procedure for assymetric damping rates}\label{SM2}

\begin{figure}[h]
\centering
\includegraphics[width=11cm]{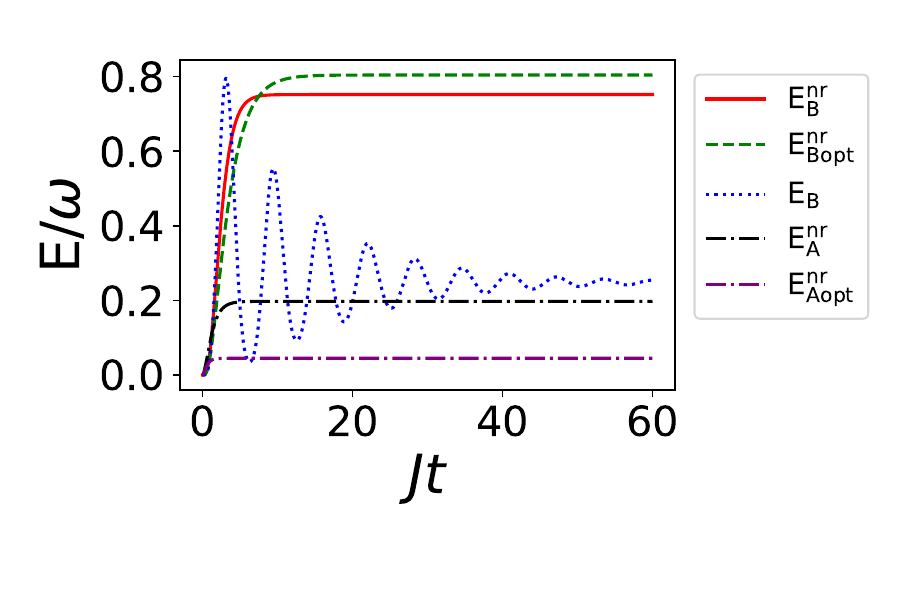}
\caption{Energies of the battery $B$ and the charger $A$ versus $Jt$. $E_B$ (dotted blue line) is the energy of the battery in reciprocal scenario and $E^{nr}_B$ (solid red line) and $E^{nr}_A$ (dash-dot black line) represent energies of the battery and the charger in nonreciprocal scenario, respectively. $E_{B, \text{opt}}^{\text{nr}}$ (green dashed line) and $E_{A, \text{opt}}^{\text{nr}}$ (purple dashed-dotted line) are the optimized energies of the battery $B$ and the charger $A$, respectively, in the presence of the nonreciprocity. Here we have $\kappa_1=0.05$, $\kappa_2=0.01$, $J=0.2$, $\Gamma=0.4$, $p_a=p_b=1$ (for nonreciprocal scenario) and the same parameters for the optimized nonreciprocal case except that $p_{a}=(\kappa_a/\kappa_b)^{1/4}$, $p_{b}=(\kappa_a/\kappa_b)^{-1/4}$.}
\label{Fig5}
\end{figure}

By reflecting the imbalance between the damping rates of the charger and the battery ($\kappa_{a}$ and $\kappa_{b}$) through properly tuning system parameters ($p_{a}$ and $p_{b}$), we will show that non-reciprocal approach can always outperform the reciprocal one, in a sense that the steady state energy in the non-reciprocal charging is greater (or equal) to any attained in non-reciprocal setting through the evolution within physically relevant regimes of parameters. To this extent, we will use the ratio parameter
\begin{equation}\label{y}
    \xi=\sqrt{\frac{\kappa_{a}}{\kappa_{b}}},
\end{equation}
and the re-scaling
\begin{equation}\label{p-rescaled}
p_{a}\rightarrow p_{a}\sqrt{x},\\ \quad
p_{b}\rightarrow \frac{p_{b}}{\sqrt{x}},
\end{equation}
such that $|\mu|=1$ is preserved, and now both $|p_{a}|=|p_{b}|=1$. The above determines the ratio between $\Gamma_{a}$ and $\Gamma_{b}$:
\begin{equation}\label{Gamma-rescaled}
\Gamma_{a}\rightarrow\Gamma x,\\ \quad
\Gamma_{b}\rightarrow\frac{\Gamma}{x}.
\end{equation}
In this notation, the energy stored in the battery for nonreciprocal case is given by 
\begin{equation}
       E_{B}^{\text{nr}}(t\rightarrow\infty) = \frac{16\omega\Gamma^2F^2}{(x\Gamma+\kappa_{a})^2 (\Gamma/x+\kappa_{a}/\xi^2)^2}.
\end{equation}
The above equation, treated as a function of $x$, achieves its maximum at
\begin{align}\label{optX}
    x_{opt}=\xi, 
\end{align}
yielding
\begin{equation}\label{optEnergyB}
    E_{B, \text{opt}}^{\text{nr}}(t\rightarrow\infty)=\frac{16\omega\Gamma^2F^2}{(\Gamma+\sqrt{\kappa_{a}\kappa_{b}})^4}=\frac{64\omega|J|^2F^2}{(2|J|+\sqrt{\kappa_{a}\kappa_{b}})^4},
\end{equation}
where nonreciprocal condition, $ J=-i\mu\frac{\Gamma}{2}$, was utilized. As seen in Fig. \ref{Fig5} the steady state optimized energy of the battery $E_{Bopt}^w$ (green dashed line) is higher than the non-optimized case (solid red line) and equal to the maximum energy of the battery attained in the first pick of the standard charging (dotted blue line). More interestingly, in the steady state, the energy remained in the charger in the optimized charging process $E_{A, \text{opt}}^{\text{nr}}(t)$ (dash-dotted purple) is less compared to the non-optimized $E_{A}^{\text{nr}}(t)$ (dash-dotted black line).

\section{Nonreciprocity advantage for all times }\label{SM3}

In the main text, the advantage of nonreciprocal scenario over reciprocal one was identified in the stationary regime. Here we prove that in the regime $\kappa_{b}<0.22 \kappa_{a}\leq |J|$ the stationary energy in the non-reciprocal scenario {\it is always better than any energy achievable by the reciprocal approach at any moment of the evolution}. We will use the ratio factor between the dissipation and the coupling
\begin{equation}
    r=\frac{\kappa_{a}}{|J|},
\end{equation}
which, defining $y:=1/\xi^2$, also implies 
\begin{equation}
    \kappa_{b} = r\cdot y\cdot |J|.
\end{equation}
In the regime $r<1$, we have an imaginary $\Delta=i\Delta_{+}$, with $\Delta_{+}=\sqrt{16|J^2|-(\kappa_{a}-\kappa_{b})^2}$. Therefore, the nonreciprocal solution is oscillatory. Now we define and analyze a time-dependent gap $D_t$ between the stationary state energy of the non-reciprocal approach, and the nonreciprocal setting:
\begin{align*}\label{Dfff}
    D_t &= E_{B, opt}^{nr}(t\rightarrow\infty)-E_{B}(t)\\
    &= 8\frac{F^2}{|J|^2}\Biggl(\frac{8}{\left(r\sqrt{y} + 2\right)^4} - \frac{\frac{2e^{-\frac{1}{2} r t(y+1)|J|}\left(|J|^2\left(8 - r^2\left(y^2+1\right)\right)\cos\left(\frac{\Delta_+ t}{2}\right)+\Delta_+ r(y+1)|J|\sin\left(\frac{\Delta_+ t}{2}\right)\right)}{\Delta_+^2}+\frac{4|J|^2\left(r^2 y+4\right)e^{-\frac{1}{2} r t(y+1)|J|}}{\Delta_+^2}}{\left(r^2 y + 4\right)^2}\\
    &- \frac{2-4e^{-\frac{1}{4} r t(y+1)|J|}\left(\frac{r(y+1)|J|\sin\left(\frac{\Delta_+ t}{4}\right)}{\Delta_+} + \cos\left(\frac{\Delta_+ t}{4}\right)\right)}{\left(r^2y + 4\right)^2}\Biggl).\numberthis
\end{align*}

\begin{figure}[h]
\centering
\includegraphics[width=7cm]{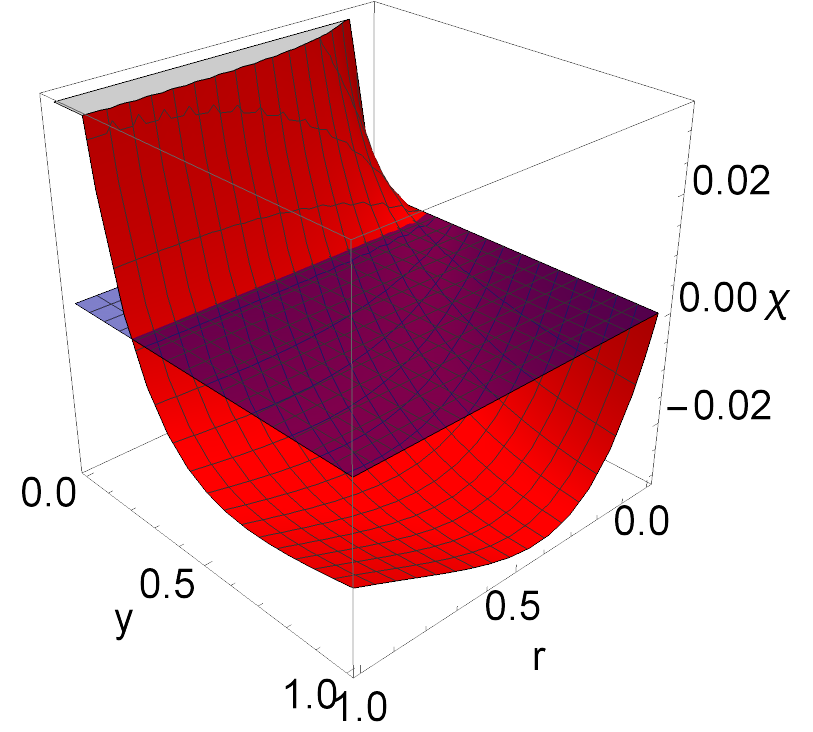}
\caption{A minimal gap $\chi$ between the non-reciprocal optimized charging in the stationary limit and finite-time dynamics of the reciprocal scenario, in the units of $\mathcal{N}$.}
\label{Chi}
\end{figure}

Below we show that the above function is positive for all $0\leq r\leq1$ and $0\leq y<0.22$. It can be shown that the function $D_t$ in Eq. (\ref{Dfff}) achieves its local minimums only at 
\begin{equation}\label{period}
    t=t^*:=\frac{(2k+1)4\pi}{\Delta_{+}},\ k=0,1,2,\dots
\end{equation}
Moreover, among these minima, the value of the function at first minimum ($k=0$) is the smallest. We see this by plugging Eq. \eqref{period} into Eq. \eqref{Dfff}:
\begin{align}
    &D_{t}\geq\chi(r,y),
\end{align}
where
\begin{align}
    \chi(r,y):=
    \mathcal{N}\left(\frac{8}{\left(r\sqrt{y}+2\right)^4}-\frac{2  \left(e^{\frac{-\pi(2k+1)r(y+1)}{16-r^2(1-y)^2}}+1\right)^2}{\left(r^2
   y+4\right)^2}\right),
\end{align}
which is a monotonically increasing function of $k$ and $\mathcal{N}\equiv 8F^2/|J|^2$. Note that $\chi$ is a continuous function of $r$ and $y$. This function is positive for $0\leq r \leq 1$ and $0\leq y < 0.22$, see Fig. \ref{Chi} with $\chi/\mathcal{N}$ plotted for $k=0$ for an illustration. Therefore, if the damping rate of the battery is at least 5 times weaker than the one of the charger, the nonreciprocal approach in the stationary limit is always advantageous over both transient and stationary limits of the reciprocal scenario.

\end{document}